\pdfoutput=1
\documentclass[cits]{myJINST}
\let\ifpdf\relax

\usepackage{subfigure}

\usepackage{url}
\usepackage{lineno}

\newcommand{\thi}{$^{\mathrm{th}}$}

\newcommand{\Lilasii}{{\sc LilasII}}
\newcommand{\ADC}{{\sc LastQDC}}
\newcommand{\Slama}{{\sc Slama}}
\newcommand{\LastRod}{{\sc LastROD}}

\newcommand{\Shaft}{{\sc Shaft}}

\newcommand{\Atlas}{ATLAS}
\newcommand{\TileCal}{TileCal}

\newcommand{\Laser}{Laser}

\newcommand{\fCs}{f_{\rm Cs}}
\newcommand{\fLas}{f_{\rm Las}}

\title{The \Laser{} calibration of the \Atlas{} Tile Calorimeter during the LHC run 1}
\author{\Atlas{} Tile Calorimeter system}
\abstract{This article describes the \Laser{} calibration system of the \Atlas{} hadronic Tile Calorimeter
that has been used during the run 1 of the LHC.
First, the stability of the system associated readout electronics is studied. It is found to be stable with
variations smaller than 0.6~\%.
Then, the method developed to compute the calibration constants, to correct for the variations of the gain of 
the calorimeter photomultipliers, is described. These constants were determined with a statistical
uncertainty of 0.3~\% and a systematic uncertainty of 0.2~\% for the central part of the calorimeter and
0.5~\% for the end-caps.
Finally, the detection and correction of timing mis-configuration of the Tile Calorimeter using the \Laser{}
system are also presented.}

\begin{document}



\begin{figure}
  \begin{center}
    \includegraphics[width=0.9\textwidth]{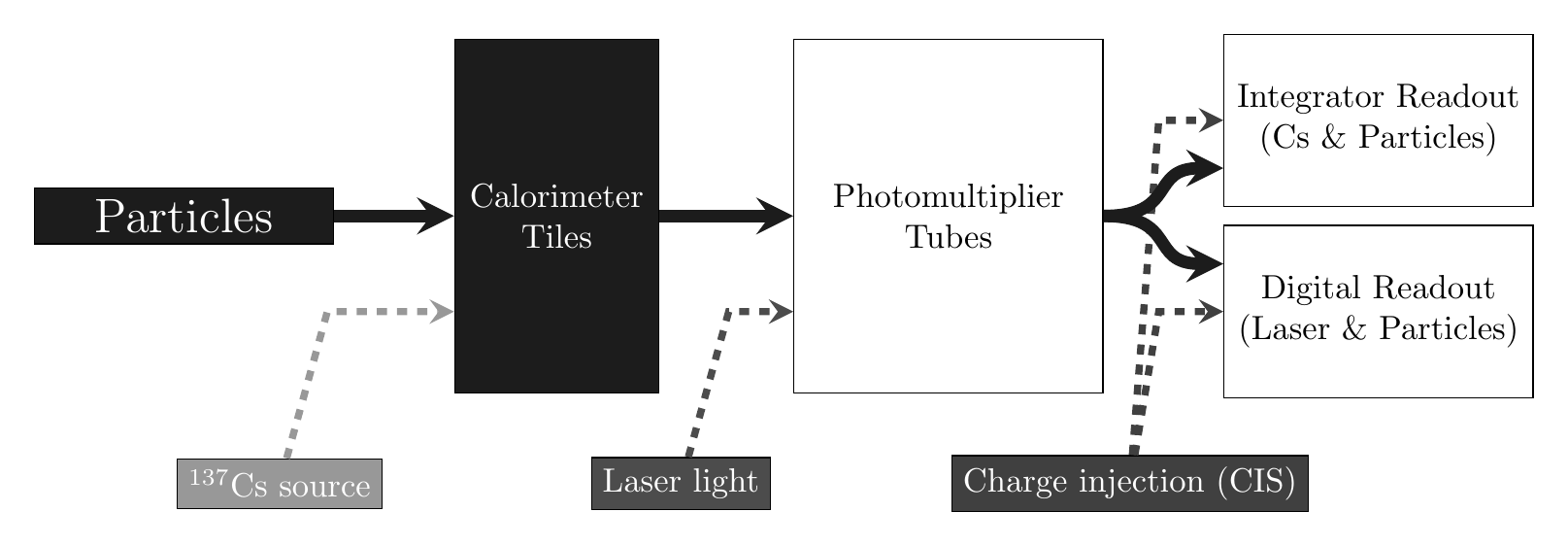}
  \end{center}
  \caption{Flow diagram of the readout signal paths of the different \TileCal{} calibration systems.}
  \label{fig:calibs}
\end{figure}

The \Atlas{} Tile Calorimeter~\cite{Aad:2008zzm,Aad:2010af} (\TileCal{}) is the central hadronic 
calorimeter of the \Atlas{} experiment at the Large Hadron Collider (LHC) at CERN.
The \TileCal{} is a sampling calorimeter whose operation is based on the
detection of scintillation light using photomultiplier tubes (PMTs). 
On average, around 30~\% of the total energy of jets from quark and gluon fragmentation
is deposited in the \TileCal{}. It therefore plays an
important role for the precise reconstruction of the kinematics of the physics event.
The control
of its stability, within 1~\%, and resolution is important for a correct jet and missing transverse
energy reconstruction in \Atlas{}\footnote{The designed energy resolution of jets is $\sigma/E=50 \%/\sqrt{E\mathrm{(GeV)}}\oplus 3 \%$ and the systematic uncertainty on the jet energy scale in \Atlas{} must be at most 1~\%. See details in~\cite{Aad:2010af}.}.
In order to obtain a precise
and stable measurement of the energy deposited in the calorimeter, it is mandatory to precisely
monitor any variation of the gain\footnote{What is really monitored is the product of the PMT photocathode
quantum efficiency by the PMT gain. In this article, it is assumed that any quantum efficiency variation
is negligible over the typical time scale of a few months and therefore the term {\em PMT gain} is used instead
of the product.} 
of the PMTs and, if needed, correct for these variations. Several
complementary hardware calibration systems have therefore been included in the \TileCal{} design
(see figure~\ref{fig:calibs}), one 
of them being the \Laser{} system described in this article.

The calorimeter is briefly described in section~\ref{sec:tile}, and the \Laser{} system as it was
operational during run 1 of the LHC is detailed in section~\ref{sec:laser}.
A major upgrade of this system was performed in 2014, but falls outside the scope of this article.
The study of the stability of the \Laser{} electronics is discussed in
section~\ref{sec:stabelec} before the description of the calibration procedure in section~\ref{sec:calib}. Timing mis-configuration detection and correction are detailed in section~\ref{sec:timing}
while monitoring of pathological calorimeter channels is briefly described in section~\ref{sec:monitor}.
Finally, conclusions are drawn in section~\ref{sec:concl}.

\section{The \Atlas{} Tile Calorimeter}\label{sec:tile}
The \Atlas{}
detector is made of a central part, called the barrel, and two
endcaps. In the barrel, the \TileCal{} is the only hadronic calorimeter
of \Atlas{}. In the endcaps, \TileCal{} constitutes the external part of the hadronic
calorimeter, the internal part using the liquid argon technology.
The part of \TileCal{} that is in the \Atlas{} barrel is called the \emph{long barrel} (LB), while
the parts that are in the endcaps are called the \emph{extended barrels} (EB). 

The Tile Calorimeter is the result of a long process of R\&D and construction.
The Technical Design Report~\cite{TDR} has been completed in 1995 and the construction of
the mechanical part ended in 2006. After a period of operation using cosmic
muons, the \TileCal{} was ready in 2009 to record the first LHC proton--proton collisions.

\subsection{Mechanical aspects}\label{sect:cells}
\begin{figure}[htp]
  \begin{center}
    \includegraphics[width=\textwidth]{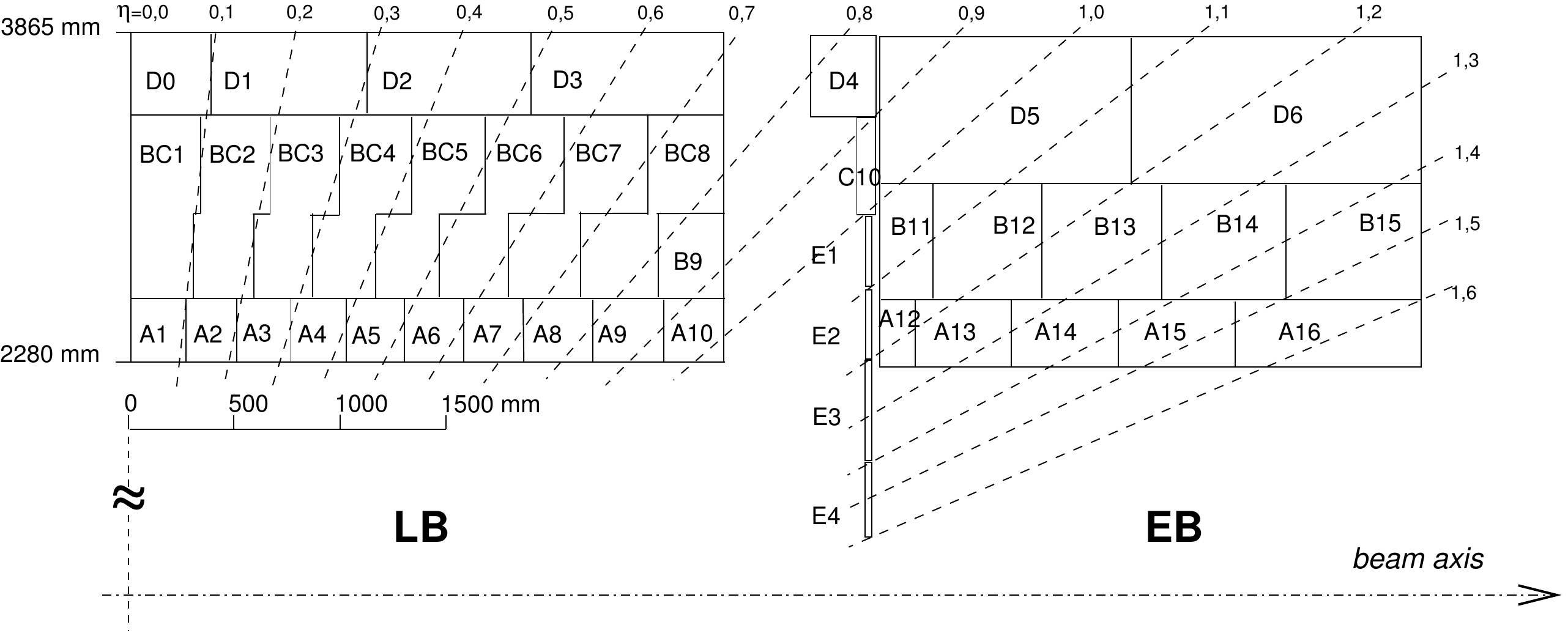}
  \end{center}
  \caption{Schematic of the cell layout in a plane parallel to the beam axis, showing
  only positive $\eta$ side (the detector being symmetric with respect to $\eta=0$).
  The three radial layers (A, BC and D) are also visible.
  Special scintillators, called gap (E1 and E2) and crack (E3 and E4) cells, are located
  between the barrel and the endcap.}
  \label{fig:cells}

  \begin{center}
    \includegraphics[width=0.5\textwidth]{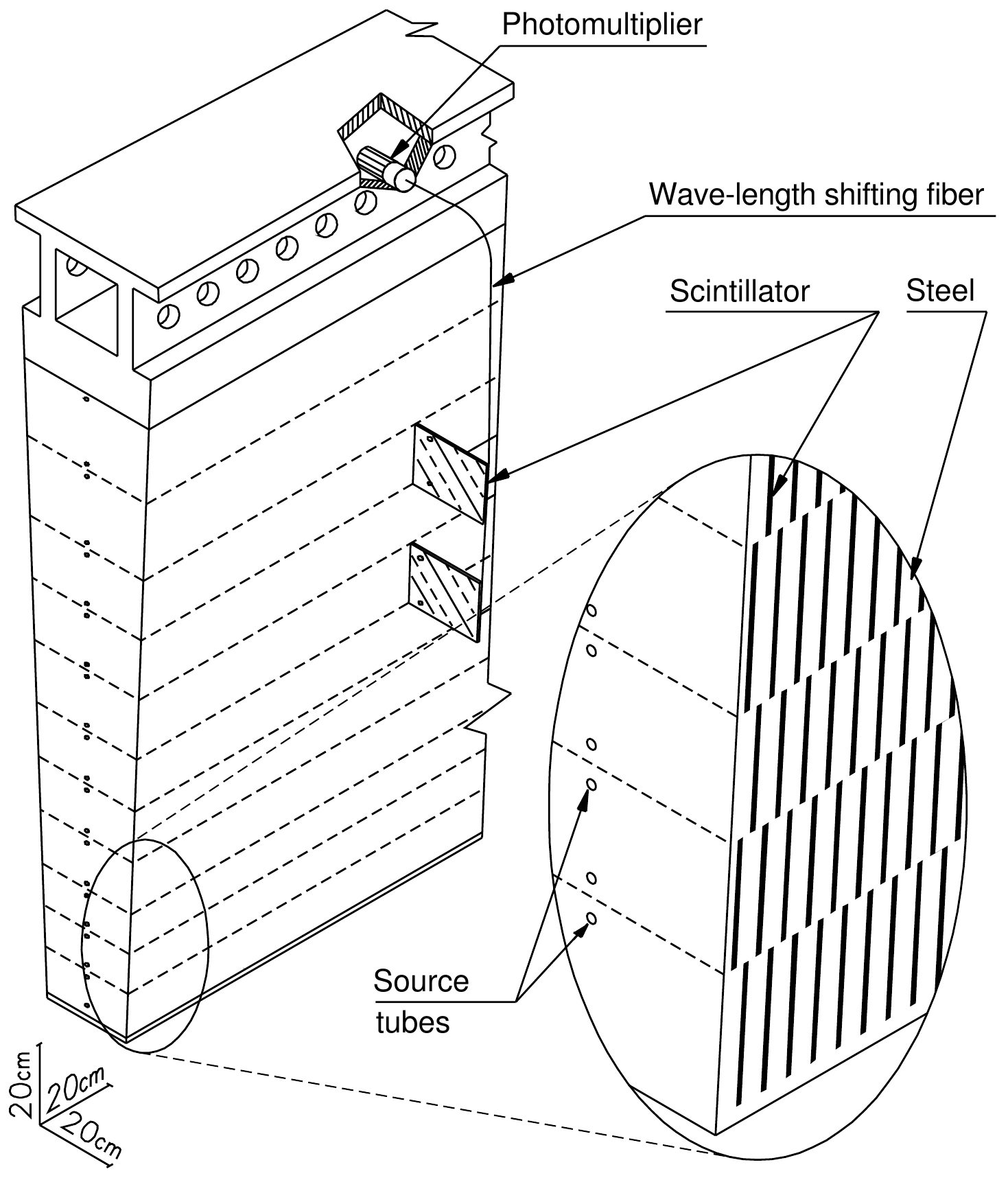}
  \end{center}
  \caption{Sketch of a \TileCal{}
  module, showing the light readout path between the scintillating tiles and the photomultiplier tubes (PMTs).}
  \label{fig:module}
\end{figure}

\TileCal{} is a non-compensating sampling calorimeter made of steel plates that act as absorber and provide
mechanical structure into which the active scintillating tiles are inserted. 
Charged particles going through the tiles produce scintillation light
that is collected
by two wavelength-shifting (WLS) fibres, on each side of a tile. These fibres
are then grouped in bundles to form cells, organised in three radial layers, as 
depicted in figure~\ref{fig:cells}, thus achieving a granularity of about 0.1 in 
$\eta$\footnote{ATLAS uses a right-handed coordinate system with its origin at the nominal interaction point (IP) in the centre of the
detector and the $z$-axis coinciding with the axis of the beam pipe. The $x$-axis points from the IP to the centre of the LHC ring,
and the $y$-axis points upward. Cylindrical coordinates ($r$,$\phi$) are used in the transverse plane, $\phi$ being the azimuthal angle around
the beam pipe. The pseudorapidity is defined in terms of the polar angle $\theta$ as $\eta =-\ln \tan({\theta/2})$.} for the layers A and BC (the closest to the collision point) and
around 0.2 in the outermost layer. Azimuthally, the detector is segmented in 64 wedge-shaped modules 
(see figure~\ref{fig:module}), thus achieving 
a granularity of $\Delta\phi=0.1$.
The E cells are non-standard calorimeter cells made of single large scintillators.
Their aim is to measure the energy of particles lost in the inactive material located in front of the
\TileCal{}. In total, there are 5182~cells.

In addition, 16 large scintillator plates are located between the barrel and the endcaps, 
the Minimum Bias Trigger Scintillators (MBTS). They are mainly used in  
the \Atlas{} low luminosity trigger. Although they are not
\TileCal{} cells, their readout is performed by dedicated PMTs of the \TileCal{} system, that were originally
planned to be connected to 8 pairs of E3 and E4 cells.

\subsection{Detector readout}\label{sect:readout}
Each fibre bundle, usually corresponding to one side of a cell, 
is read out by a photomultiplier tube:
each standard cell is thus read out by two PMTs, the E cells being read out by a single PMT.
Therefore, there are 9852~PMTs in total. The
electric pulses generated by the PMTs are shaped~\cite{Anderson:2005ym} and 
digitised~\cite{Berglund:2008zz} at 40~MHz with two 
different gains, with a ratio of 64, in order to achieve a good precision in a wide energy range. 
These samples are then stored in a
pipeline memory until the level-1 trigger decision is taken 
(\Atlas{} has a three-level trigger system, the first
level giving a decision in 2.5~$\mu$s during which the data are kept in the front-end electronics). 
If the decision
is positive, seven samples, in time with the signal and with appropriate gain 
giving the best precision on the
pulse amplitude, are sent to the off-detector electronics (Read Out Drivers or RODs~\cite{Poveda:2007zz}). 
The amplitude
of the signal is reconstructed as the weighted linear combination of the digitised
signal samples, using an optimal filtering method~\cite{Cleland:2002rya,Valero:2009wda}.

In parallel, the output
of each PMT is also integrated over approximately 14~ms
using an analog integration system, during Cesium calibrations (see below and figure~\ref{fig:calibs}) 
and also to measure the PMT current induced by minimum bias proton--proton collisions.

\subsection{Hardware calibration systems}
Three different and complementary hardware calibration systems have been integrated
in the \TileCal{} design: the Cesium system~\cite{Starchenko:2002ju}, the \Laser{} system 
and the Charge Injection System (CIS)~\cite{Anderson:2005ym} (see figure~\ref{fig:calibs}).

About once or twice a month, a radioactive $^{137}$Cs source scans most\footnote{For mechanical reasons, the special cells E3 and E4 cannot be calibrated by the Cesium source, see Fig.~\ref{fig:cells} and Section~\ref{sect:det_consts} for more details.}
\TileCal{} cells, measuring the response of the tiles, including the WLS fibres
and the PMTs, to a known amount of deposited energy. For this Cesium calibration, the
analog integrator is used instead of the digital readout: it is therefore not sensitive
to potential variations of the gain of the readout electronics used for physics.

The \Laser{} calibration is performed about twice a week and 
allows monitoring the stability of the
PMT gain between two Cesium calibrations by illuminating all PMTs with a \Laser{} pulse
of a known intensity. For this calibration, the same readout electronics as for
physics is used and is therefore also monitored.
This calibration is described in detail in the next sections. 

During the CIS calibrations, performed about twice a week, a known electric charge 
is injected into the readout electronics chain, simulating a PMT output pulse. 
It is used to measure the conversion factor
from ADC-counts to pico-Coulomb (pC) and also to
monitor the linearity of the ADCs.

\begin{figure}
  \begin{center}
    \includegraphics[width=\textwidth]{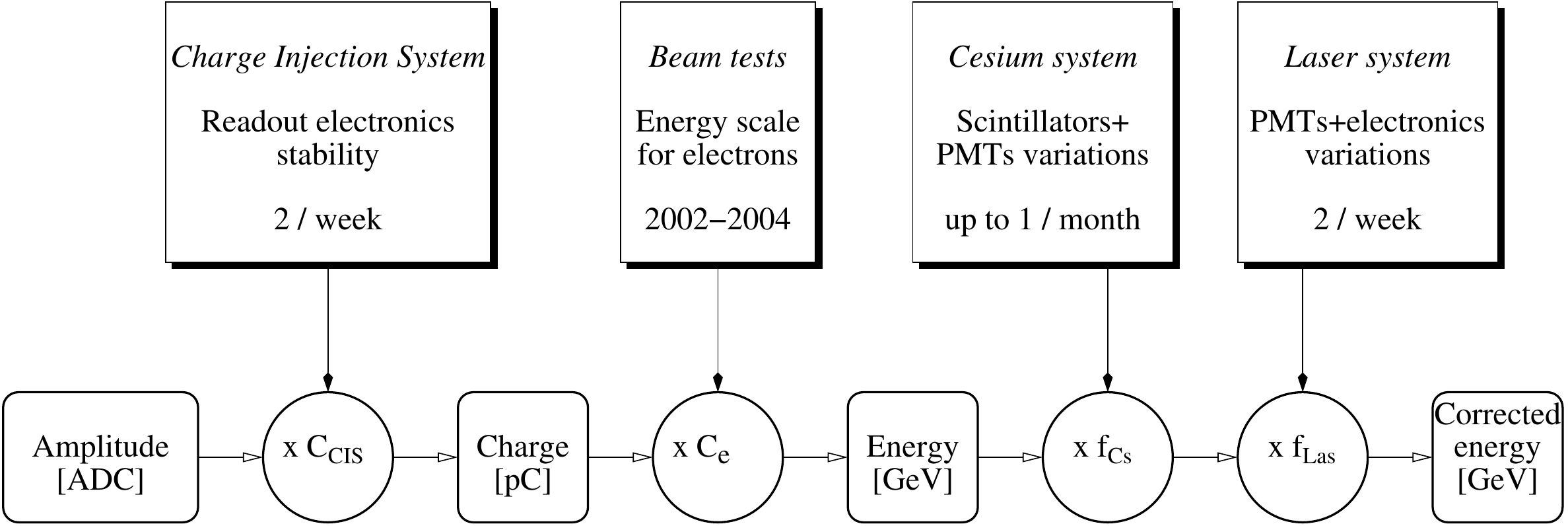}
  \end{center}
  \caption{Diagram of the contributions of the hardware calibration systems to the energy
    reconstruction.}
  \label{fig:calib_sketch}
\end{figure}

\subsection{Energy reconstruction}
In order to measure correctly the energy, several effects must be taken into account, using the
calibration systems described previously as well as results from tests using special beams of particles (see figure~\ref{fig:calib_sketch}).
Hence, for each PMT, the energy is reconstructed as
\begin{equation}\label{eq:E}
  E=A_{\rm opt}\times C_{\rm CIS}\times C_e \times \fCs \times \fLas
\end{equation}
where $A_{\rm opt}$ is the amplitude in ADC counts computed by the optimal
filtering method, $C_{\rm CIS}$ is the ADC$\rightarrow$pC conversion factor
measured by the CIS, $C_e$ is the pC$\rightarrow$GeV conversion factor 
defining the energy scale as measured with electron beams in past beam tests~\cite{Adragna2009362} and
$\fCs$ and $\fLas$ are correction factors extracted from the 
Cesium and \Laser{} calibrations. The Cesium calibration is able to correct the
residual non-uniformities after the gain equalisation of all channels and thus to preserve the
energy scale of the calorimeter that was determined during the beam tests.
The Laser calibration allows keeping stable this energy scale between two Cesium scans and is
therefore relative to the Laser calibration immediately consecutive to the last Cesium calibration.
The cell energy is the sum of the energies from the two PMTs connected
to this cell. 

The computation of the \Laser{} calibration constants $\fLas$ will be described in the next
sections.

\section{The \Laser{} calibration system}\label{sec:laser}
The first prototype of the \Laser{} system has been built in 1993 and the first tests with a prototype
module of the calorimeter were performed in October 1994~\cite{Ajaltouni:683494}. A second prototype
was then built and tested in 1997~\cite{Grenier:683674}. After several
improvements, a first version of the system has been used during the qualification of the \TileCal{}
modules using particle beams~\cite{Adragna2009362}. Finally, the system described in the following
paragraphs was installed
in its definitive location in 2008, in time for the calibration of \TileCal{} before the first
LHC collisions.
After several years of running, the current \Laser{} system has been upgraded in October 2014 for the LHC run 2,
starting in 2015. This new system includes a new \Laser{} source, new optical components
and new electronic boards, and is not described in this article.

The \Laser{} system contains of course a \Laser{} source but also other components.
Because the actual intensity of the pulses has 
variations of approximately 5~\%, the system includes dedicated photodiodes to precisely measure the 
intensity of each pulse. It also includes a tunable attenuation system to cover the whole 
\TileCal{} PMTs dynamic range (100~MeV to 1.5~TeV per PMT).

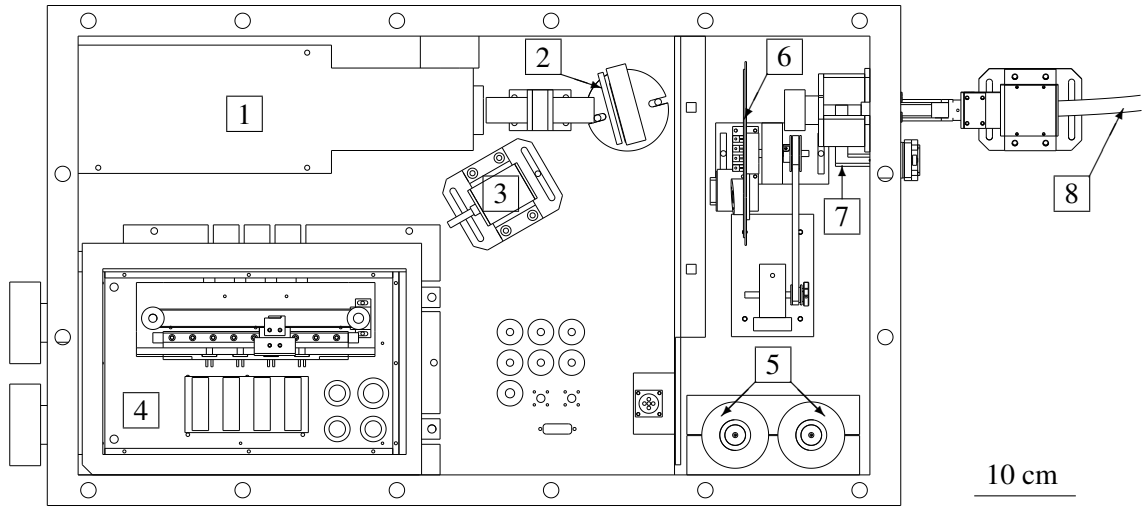
\begin{figure}[t]
  \begin{center}
    \setlength{\unitlength}{0.1\textwidth}
    \begin{picture}(10,5)
      \put(0,0){\includegraphics[width=\textwidth]{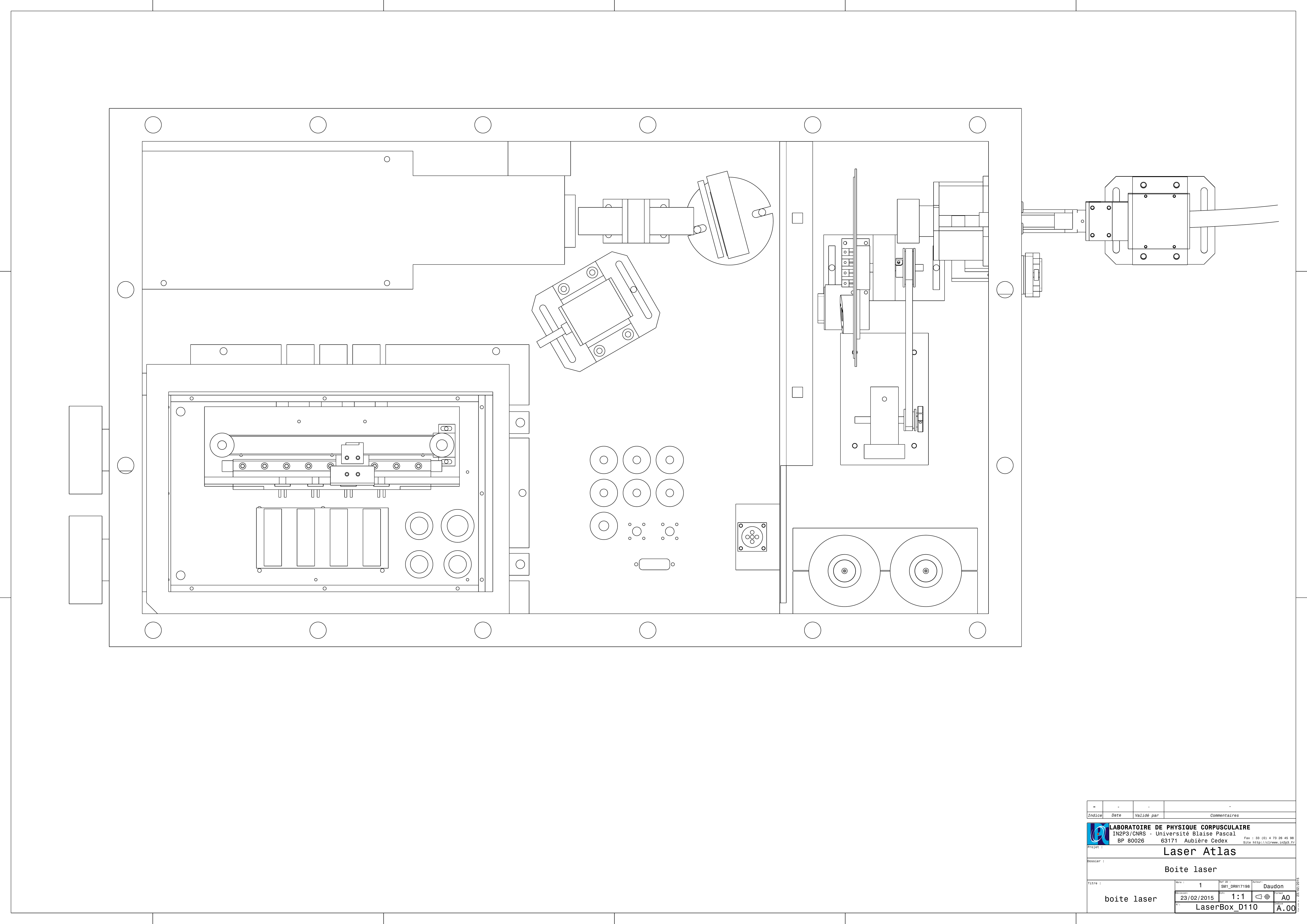}}
      \put(2,3.7){\framebox(0.3,0.3){1}}
      \put(4.6,4.2){\framebox(0.3,0.3){2}}
      \put(4.9,4.2){\vector(3,-1){0.36}}
      \put(4.23,3){\framebox(0.3,0.3){3}}
      \put(1.1,1.1){\framebox(0.3,0.3){4}}
      \put(6.6,1.5){\framebox(0.3,0.3){5}}
      \put(6.6,1.5){\vector(-1,-1){0.3}}
      \put(6.9,1.5){\vector(1,-1){0.3}}
      \put(6.7,4.2){\framebox(0.3,0.3){6}}
      \put(6.7,4.2){\vector(-1,-2){0.2}}
      \put(7.2,2.8){\framebox(0.3,0.3){7}}
      \put(7.35,3.1){\vector(0,1){0.3}}
      \put(9.2,3){\framebox(0.3,0.3){8}}
      \put(9.5,3.3){\vector(1,2){0.3}}
      \put(8.5,0.5){\line(1,0){0.884}}
      \put(8.6,0.6){10 cm}
    \end{picture}  
  \end{center}
  \caption{Sketch of the \Laser{} box, showing the \Laser{} head (1), the semi-reflecting mirror (2),
    the light mixer (3) that distributes the light to the photodiodes box (4) and the two PMTs (5),
    the filter wheel (6), the shutter (7) and the output liquid light guide (8).}
  \label{fig:boxes}
\end{figure}

The \Laser{} system consists of a \Laser{} pump, the so-called \Laser{} box, 
a light distribution system and several electronic components in a VME crate to control
and monitor the system. The \Laser{} box (see figure~\ref{fig:boxes}) 
contains the \Laser{} head, two photomultipliers, the
so-called photodiodes box, a semi-reflecting mirror, a filter wheel and a shutter. 
The photodiodes box contains four PIN
photodiodes from Hamamatsu~\cite{PhD} with their amplifiers and shapers. These components are
located in the USA15 cavern that hosts the back-end electronics, and that is 
close to the main \Atlas{} cavern.

The \Laser{} system can be operated in four different modes, divided in two categories: three 
internal calibration modes and one \Laser{} mode.
\paragraph{Internal calibration modes:} the aim of these modes
is to calibrate and monitor the \Laser{} system electronics, without any \Laser{} light emission. 
There are three different calibration modes:
\begin{itemize}
\item two to measure the characteristics of the electronics: linearity (Linearity mode), noise and pedestal
level (Pedestals mode);
\item one to calibrate the photodiodes, with an embedded radioactive source (Alpha mode).
\end{itemize}
\paragraph{\Laser{} mode:} this mode represents the main operation mode, in order to calibrate the \TileCal{} PMTs using
\Laser{} light.

\subsection{Choice of the light source}
In order to synchronise all the \TileCal{} channels, it is mandatory to be able to send simultaneously
a single light pulse to the 9852 PMTs, which requires a powerful light source such as a \Laser{}. 
Pulse shape requirements ($\sim$10~ns pulse width) has driven our choice toward a commercial Q-switched DPSS (Diode-Pumped Solid State) \Laser{} manufactured by SPECTRA-PHYSICS. This is a frequency-doubled infrared \Laser{} emitting a 532~nm green light beam, this wavelength being close to the one of the light coming out of the WLS fibres (480~nm). It delivers 10 to 15~ns pulses of a few $\rm{\mu J}$ maximum energy, which is sufficient to saturate all \TileCal{} channels, and thus to test their full dynamic range.
Moreover, the pulse shape is sufficiently similar to the shape of physiscs signals, so that the optimal
filtering method does not have to be adapted. The delay between triggering and actual light emission is of the order of 1.2~$\mu$s, which depends on the pulse energy in a window of about 60~ns; and a jitter of 25~ns due to the internal 40~MHz clock of the \Laser{} source electronics.

\subsection{\Laser{} light path}
The light emitted by the \Laser{} head passes first through a semi-reflecting mirror (commercial metallic 
neutral density filter). About 10~\% of the light is redirected towards a specific 
calibration system in order to measure precisely the amount of light that has been emitted, with a
photodiode, and also
to precisely measure the time when the pulse was emitted with two PMTs. 

\begin{figure}[htp]
  \begin{center}
    \includegraphics[width=0.7\textwidth]{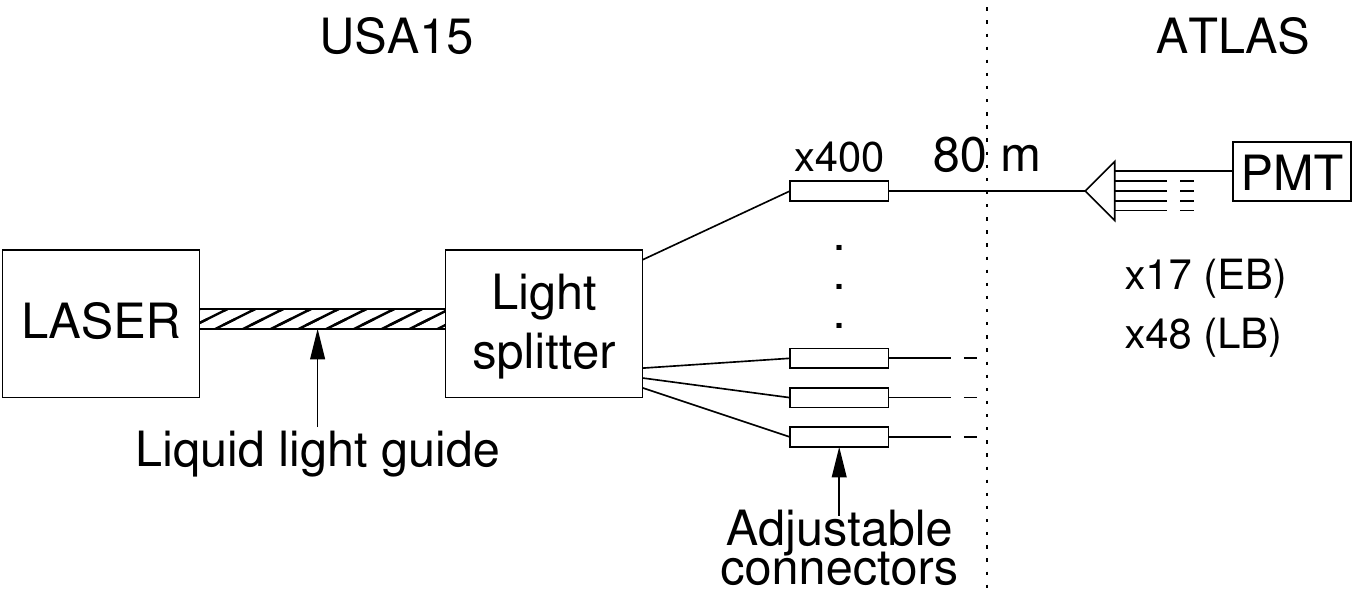}
  \end{center}
  \caption{Sketch of the light splitting scheme. The main light splitter distributes the light to
    400 optical fibres, 256 for the two extended barrels, 128 for the long barrel and 16 spares.
    In each module, the secondary light splitter distributes the light to 17 (EB) or 48 (LB) fibres,
    connected to 16 (EB) or 45 (LB) photomultipliers.}
  \label{fig:lightdistr}

  \begin{center}
    \setlength{\unitlength}{0.1\textwidth}
    \begin{picture}(10,5)
      \put(0,-1){\includegraphics[width=\textwidth]{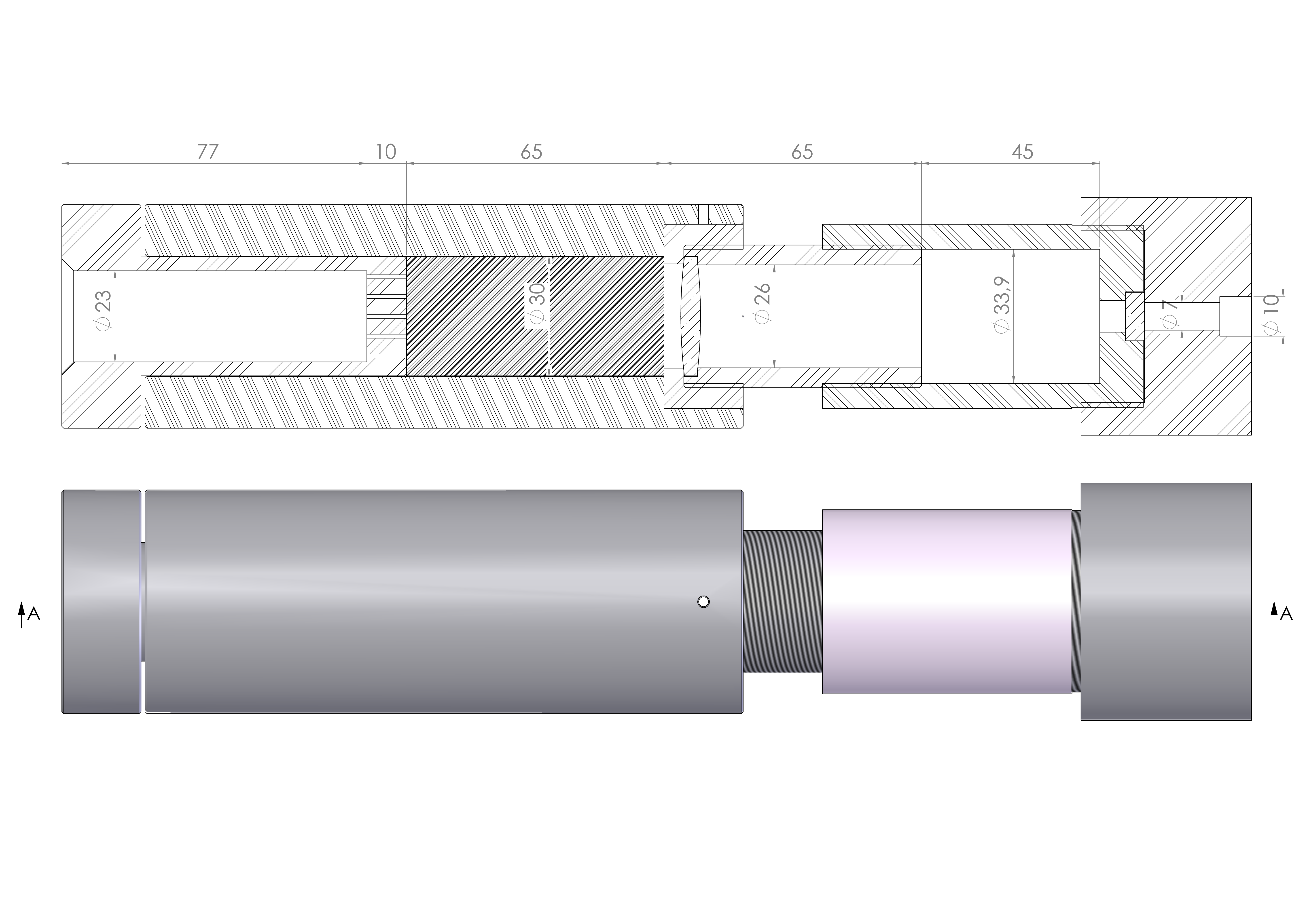}}
      \put(6.6,0.1){\line(1,0){1.47}}
      \put(7,0.2){5 cm}
    \end{picture}  
  \end{center}
  \caption{Sketch of the beam expander and light splitter. The light enters from the right hand-side of
    the figure.}
  \label{fig:splitter}
\end{figure}

The light that passes through the semi-reflecting mirror then goes through a rotating filter wheel, 
which is used to attenuate the outgoing light. This wheel contains seven metallic neutral density filters, 
each one applying a specific attenuation factor (from 3 to 1000). 
The filter wheel contains also an empty slot in order to reach the maximum intensity.

Downstream of the filter wheel, a shutter allows controlling whether the light is sent to \TileCal{} or not. 
When this
shutter is open, the light enters a 1~m long light guide with a 5~mm diameter liquid core
that acts as a first beam expander (see figure~\ref{fig:lightdistr}).
The liquid light guide has been chosen for its large acceptance cone, high transmittance and
very good resistance to powerfull light pulses.
At the output of the liquid light guide, the light enters
in the beam expander and light splitter~\cite{Alves:466490} (see figure~\ref{fig:splitter}), where
the \Laser{} beam is first enlarged by means of a
divergent lens followed by a convergent one in order to obtain a parallel light beam. 
In order to avoid any speckle effects, the beam then goes into a coherence-killing light-mixer, made
of a parallelepipedic PMMA block. At the output of this block, the light is transmitted to a bundle of
400 clear optical fibres (Mitsubishi ESKA GH4001).

In order to further improve the splitting uniformity, the amount of light transmitted to the 400 secondary 
fibres can be varied with adjustable distance optical connectors~\cite{Aitamar:683555}. 
From these 400 outputs, 384 of them are connected to 80~m long clear fibres of similar model, 
that bring the light from the USA15 cavern to the \TileCal{} modules. Inside a module, each
clear fibre is connected to 17 (EB) or 48 (LB) fibres of the same type but with lengths varying
between 0.5 and 6~m; each of these fibres reaching one PMT. The light splitting is performed by the
mean of an empty anodized aluminium tube: the incoming beam expands and reflects on the inner wall 
of the tube before reaching the bundle of output fibres. These tubes are 26~mm (respectively 40~mm) long
with an internal diameter of 6~mm (10~mm) in the EB (LB).
In total, each endcap module is fed by 
two 80~m fibres (16 PMTs per fibre) and each barrel module is fed by two fibres (45 PMTs per fibre). 
It should be noted that the light splitters located in the modules can not be tuned.

Connected to the adjustable connectors, 
three additional fibres go back inside the \Laser{} box in order to measure
the intensity of the light pulses downstream of the filter wheel. Due
to the large range of available attenuations (from 1 to 1000), the light sent to these three 
photodiodes has been tuned in order to always have a measurable signal in at least one of them.

\subsection{Radioactive source}
The four photodiodes are located in a special box in which temperature and humidity levels are monitored and
controlled by Peltier elements and a dry air flow. This box also contains a 16~mm diameter radioactive
source of $^{241}$Am, releasing mostly $\alpha$ particles of 5.6~MeV with an activity of 3.7~kBq. 
This source provides an absolute calibration (independent from the 
\Laser{} source) of each photodiode. In order to calibrate all the diodes, the source can be 
moved along them into the box.

\subsection{Electronics}
\begin{figure}[t]
\centering
\includegraphics[width=0.9\columnwidth]{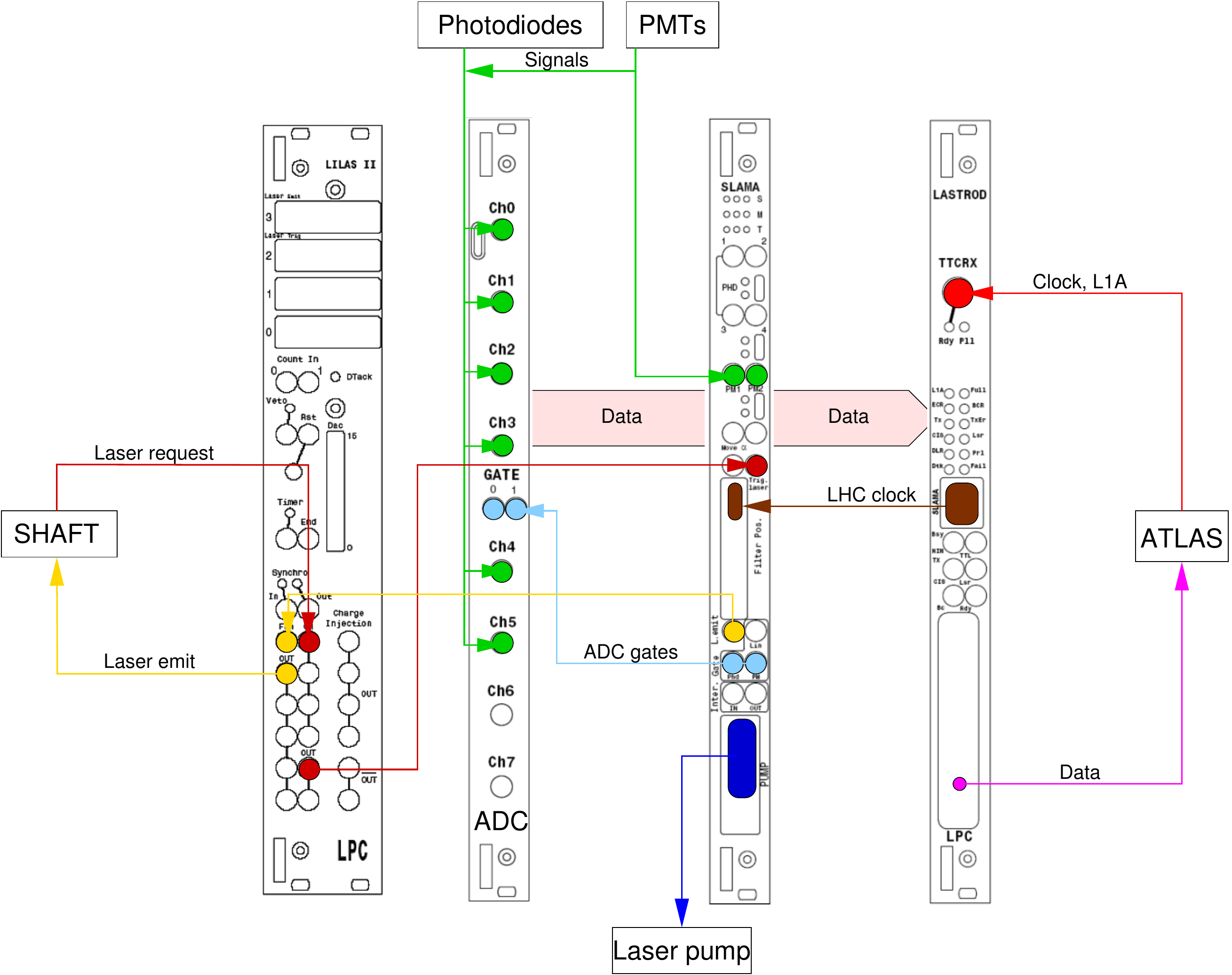}
\caption{\Laser{} system electronic boards involved in the \Laser{} mode. The interactions between the
various components are also shown. The board labelled ADC is the \ADC{} board.}
\label{fig:hard2}
\end{figure}

The VME 6U crate contains a VME processor, on which the software interface between the system
and the \Atlas{} data acquisition system (DAQ) is running, and several custom-made electronic boards.
The main boards that are needed for the data acquisition are described in the following paragraphs and
can be seen on figure~\ref{fig:hard2}.

The \ADC{} board is an eight channels charge Analog to Digital Converter, for the digitisation of
the signal coming from the four photodiodes and the two photomultipliers. 
It has a resolution of 11~bits with a maximum range of 200~pC.

The \Lilasii{} board provides two distinct functions. The first function is a charge injection system used
in the Linearity mode, with a 16-bit Digital to Analog Converter (DAC) producing a maximum
output of 10~V that is injected in a 1~pF capacitor. The second function is 
an interface between the \TileCal{} calibration requests system (\Shaft\footnote{The \Shaft{} board is
also a VME board but it is not located in the \Laser{} VME crate, as it is not only used by the 
\Laser{} system but also by the CIS and the minimum bias current measurement system. This is the board
that defines when the various systems must be triggered.}) 
and \Slama, containing mostly
counters to know precisely how many requests were sent by \Shaft{} and by \Slama.

The \Slama{} board is the main component of the system, producing signals to control other
components like the \ADC{}, \Lilasii{} or the \Laser{} pump. It contains three FPGAs (Altera ACEX
EP1K300QC208): one is devoted to the VME communication, the second one to the communication with
\LastRod{} via a dedicated bus, and the last one to the management of the \Laser{} pump and the production
of the various electronic signals that are described in the next sections. This board contains also
two 15-bit Time to Digital Converters (TDC) with a resolution of 250~ps. 
The internal clock of \Slama{} has a frequency of 80~MHz, either from an internal oscillator or
provided by \LastRod{} (see hereafter).

The \LastRod{} board is the dedicated Read Out Driver (ROD) of the \Laser{} system, included in the
\Atlas{} trigger and data acquisition system. It receives timing and trigger information from the
standard \Atlas{} TTC systems, via an optical fibre connected to a TTCrq mezzanine board~\cite{TTCrq}. 
This mezzanine
includes a QPLL that provides 40~MHz and 80~MHz clocks synchronised to the LHC clock. \LastRod{}
also sends the data to the \Atlas{} data acquisition computers (ROS), using another mezzanine board named HOLA~\cite{HOLA}, that contains a
2.5~Gbps optical link. The various functions of \LastRod{} are implemented in four FPGAs (Altera
Cyclone I EP1C12Q240C6N): the first one manages the communication with the TTCrq mezzanine,
the second one the communication with the HOLA board, the third one is the VME decoder and the last one
is devoted to the communication with the \Slama{} and \ADC{} boards via the dedicated bus.

Two other custom VME boards are needed to control the movements of the filter wheel and the
radioactive source, and to monitor the low and high voltages, temperatures and humidity levels.

\subsection{Operating in internal calibration modes}
Operating the \Laser{} system in one of the three internal calibration modes requires only a sub-set of the
\Laser{} system hardware.

\paragraph{Pedestals mode}
The aim of this mode is to record a high number of events without any energy in the
photodiodes and the photomultipliers (neither from the \Laser{} nor from the radioactive source). 
\Slama{} generates a random gate for the \ADC{}.

\begin{figure}[t]
  \begin{minipage}[t]{0.46\textwidth}
    \includegraphics[width=\columnwidth]{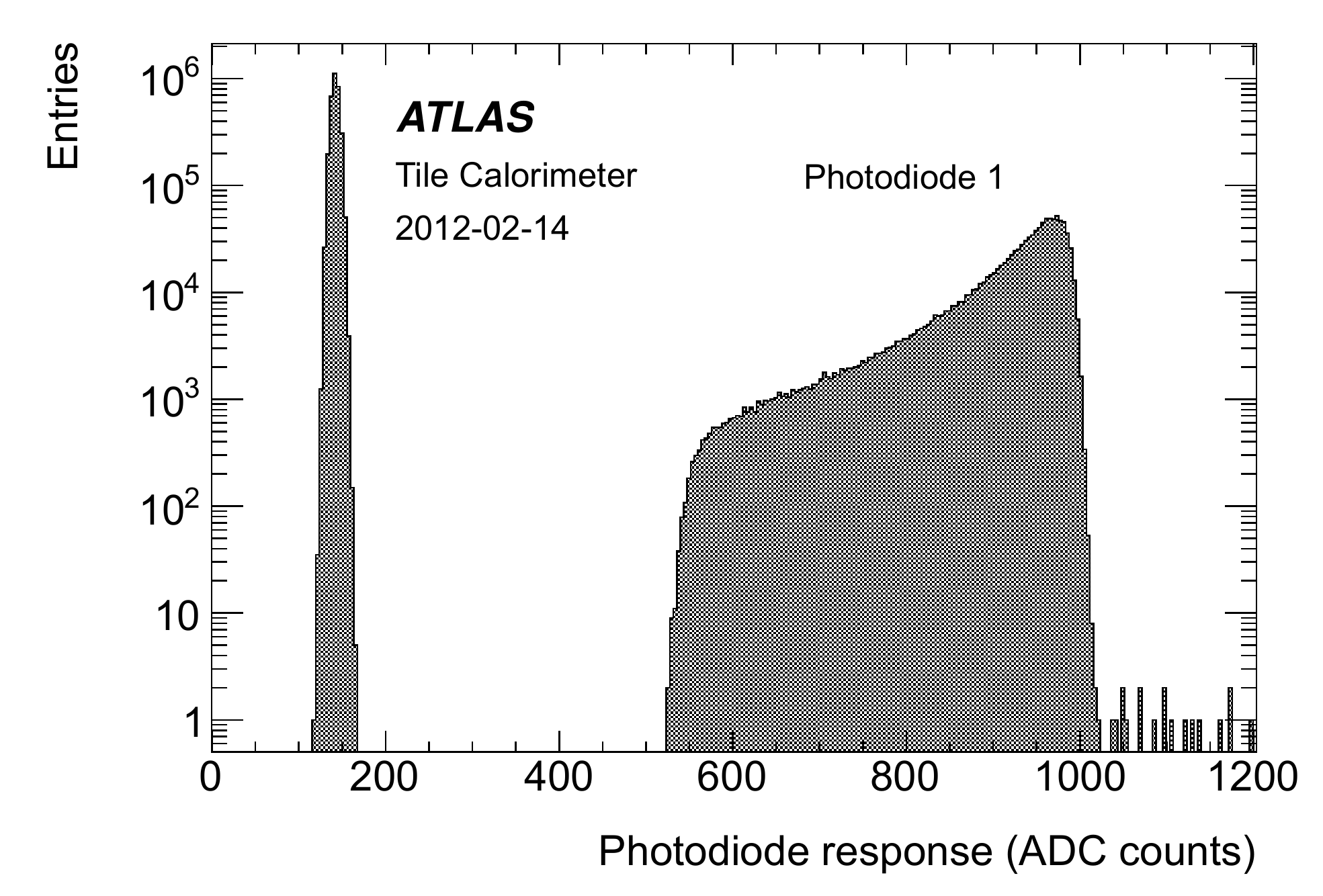}
    \caption{Example of distribution of the photodiode response in the Alpha mode. Data around ADC=170
      were taken while the radioactive source was in front of the other photodiodes, data between 500 and 1000
      correspond to the $\alpha$ spectrum.}\label{fig:calib_alpha}
  \end{minipage}
  \hspace{0.06\textwidth}
  \begin{minipage}[t]{0.46\textwidth}
    \includegraphics[width=\columnwidth]{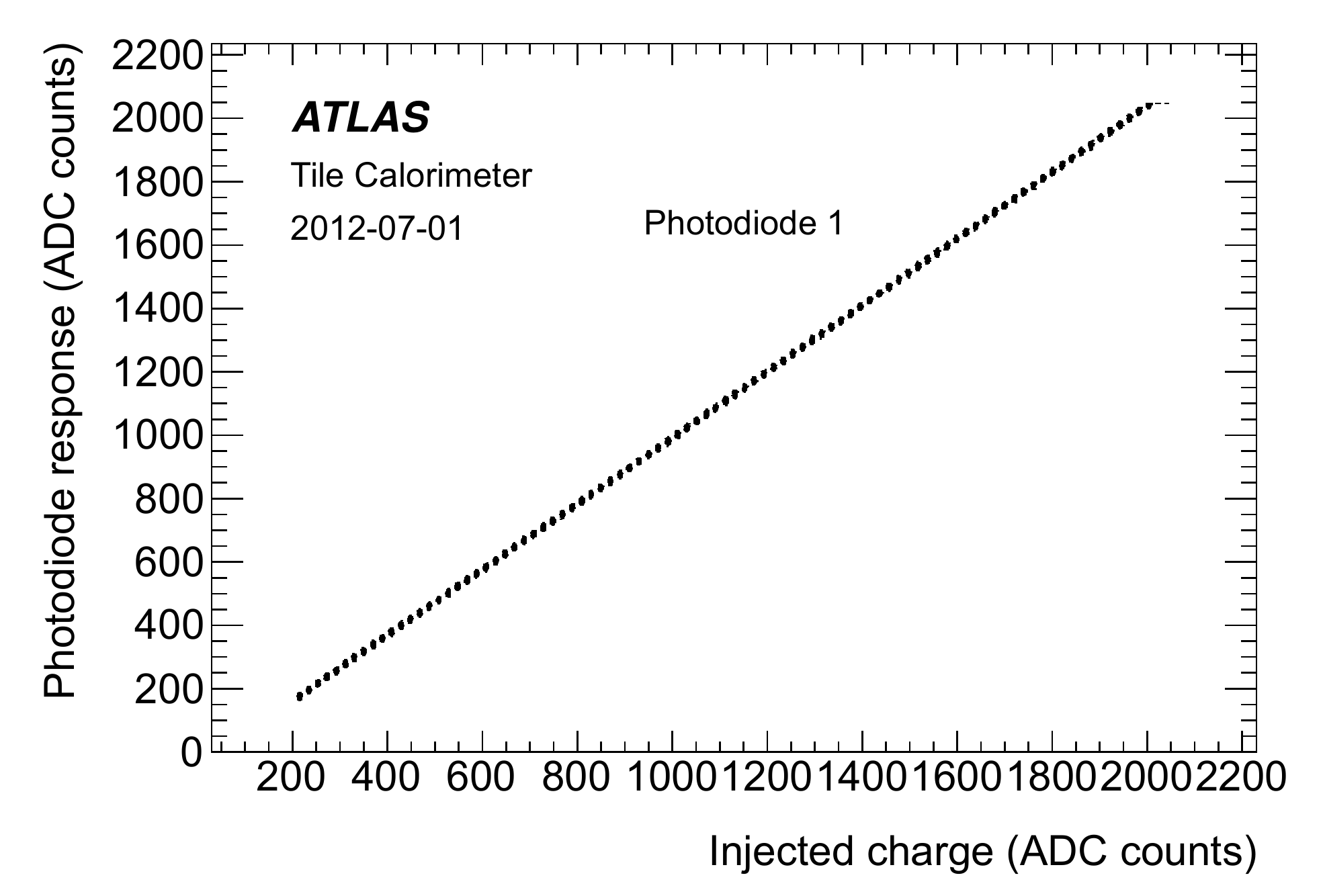}
    \caption{Example of photodiode response as a function of the measured injected charge in the Linearity mode. The saturation at high value of the photodiode response is an effect due to the ADC, not to the photodiode itself.}\label{fig:calib_lin}
  \end{minipage}
\end{figure}

\paragraph{Alpha mode}
In this mode, the $\alpha$ radioactive source is moved in the photodiodes box, in front of each
photodiode. The signals coming out of the four photodiodes are fed into \Slama{}, where
they generate an internal trigger if at least one of them is above a given threshold. This 
internal trigger generates the gate needed by the \ADC{} to digitise the delayed photodiodes
signals: for each event, only one photodiode contains energy from the $\alpha$ radiation (see
figure~\ref{fig:calib_alpha}). In very rare cases, two $\alpha$ particles hit the photodiode
within the same digitisation gate (as can be seen above ADC$\simeq$1000 in 
figure~\ref{fig:calib_alpha}) but the effect is negligible. The large dispersion of the $\alpha$ spectrum
is due to the absence of collimation of the $\alpha$ particles, resulting in a wide range of incidence 
angles\footnote{The source and the photodiode have surfaces of approximately 200~mm$^2$ and they are separated
by about 1~mm. Particles emitted perpendicular to the source travel a very short distance in the air gap and
loose a very small amount of energy, while particles emitted almost parallel to the source travel much longer
distances and therefore loose much more energy.}
for the particles hitting the photodiode.

\paragraph{Linearity mode}
This mode allows measuring the linearity of the photodiodes electronics (amplifier, shaper and ADC): 
a tunable electric charge is injected in this electronics (downstream of
the photodiodes), the response of the photodiodes electronics and the injected charge are recorded (see
figure~\ref{fig:calib_lin}).


\subsection{Operating in \Laser{} mode}\label{sect:laser_mode}
The \Laser{} system is operated in the \Laser{} mode for two data-taking situations:
\begin{itemize}
\item dedicated calibration runs, when \TileCal{} is operated independently of the rest of \Atlas{}
and all events are \Laser{} calibration events;
\item standard physics runs, when \TileCal{} is synchronised with the other \Atlas{} sub-detectors.
In this configuration, the \Laser{} pulses are emitted only during a dedicated period of the LHC
orbit, in which it is ensured there are no collisions (see figure~12.2 of 
reference~\cite{Evans:2008zzb}), in order to separate physics events
from calibration events.
\end{itemize}
The main difficulty in this mode is the timing. First, the \Laser{} system
must be synchronised with the other \TileCal{} components, so that the events recorded in \TileCal{} 
correspond to the events when the \Laser{} pulse is emitted. Second, during standard physics runs,
the \Laser{} pulses must not be emitted during proton--proton collisions. In order to achieve these
two goals, the \Laser{} system is operated as a slave of the \Shaft{} board.
The frequency at which the \Laser{} pulses are emitted is about 1~kHz in the calibration runs and
1~Hz in the physics runs.

Operating in \Laser{} mode requires most of the components, in particular the interfaces with
\Atlas{}. The various boards located in the \Laser{} VME crate that are needed in this mode can
be seen in figure~\ref{fig:hard2}, together with their interactions.

\begin{figure}[t]
\centering
\includegraphics[width=\columnwidth]{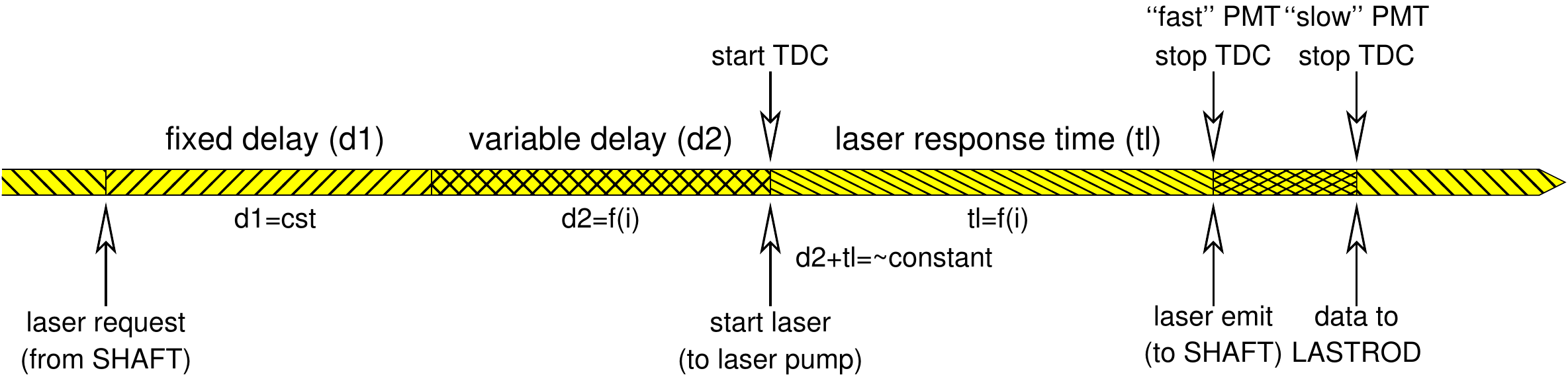}
\caption{Sequence of signals to trigger a \Laser{} pulse. The time (tl) between the reception of the
trigger signal by the \Laser{} pump and the pulse emission depends on the pulse intensity.
The programmable delay (d2) is chosen as a function of the intensity in order to keep d2+tl
constant.}
\label{fig:timing}
\end{figure}

The \Shaft{} board is programmed to send a request to the \Laser{} system at a fixed time with respect
to the beginning of the LHC orbit. Depending on the required frequency, this signal is only sent
during selected orbits. When \Slama{} receives this signal, it triggers a \Laser{} pulse by the mean
of a signal
sent to the \Laser{} pump together with a DAC level to set the required \Laser{} intensity.
However, a small delay occurs between the pump triggering and the emission of the \Laser{} pulse,
and this delay is dependent on the pulse intensity. In order to compensate
for this variable delay, \Slama{} sends the trigger signal to the \Laser{} pump only after a 
programmable time, by steps of 12.5~ns, as explained in figure~\ref{fig:timing}.


The gates needed by the \ADC{} to digitise the photodiodes response are then
produced by \Slama{} from the response of one of the \Laser{} box 
photomultipliers. Using its TDC and the signal from these photomultipliers, \Slama{}
also measures the time needed by the \Laser{} pulse to be emitted.
In order to get relevant time values, the internal 80~MHz
\Slama{} clock is the one that is produced by \LastRod{}, which is synchronised with the LHC
clock.



Once \Slama{} has detected that the pulse is correctly emitted, it generates a signal back to 
\Shaft{} that then sends a \Laser{} calibration request to the \Atlas{} central trigger processor.
This request is sent to \Shaft{} when the pulse is emitted in order to synchronise the pulse
emission and the readout of \TileCal{}. Then,
a level-1 trigger (L1A) signal associated to this \Laser{} event is distributed to the 
whole \Atlas{} detector,
together with the \TileCal{} \Laser{} calibration trigger type (TT).

Finally, when a L1A is received by \LastRod{}, associated with the \Laser{} TT,
it sends to the \Atlas{} DAQ the amplitude of the \Laser{} light, measured by the photodiodes, 
the time measured by the TDC, the average and RMS of the reponses measured during the last Pedestals 
and Alpha runs, as well as the state of the system (filter wheel position, shutter state,
temperature and humidity in the \Laser{} box, status of the power supplies). All this information is
then available for offline analysis.


\section{Stability of the electronics}\label{sec:stabelec}

Before any use of the \Laser{} system to calibrate the \TileCal{}, it is necessary to monitor the
behaviour of its readout electronics and in particular the photodiodes. This is done in two steps:
the first step is a measurement of the electronics characteristics when no energy is deposited in the
photodiodes (no \Laser{} light pulse nor $\alpha$ particle), 
the second step is a calibration of the photodiodes response using the 
radioactive source. These two steps are performed using the Pedestals and Alpha internal calibration
modes, that are taken regularly.

\subsection{Characteristics of the electronics}
Each readout channel can be characterised by two numbers: its noise and its pedestal.
The noise can be measured as the
RMS of the distribution of the responses when no energy is recorded in the channel. 
The pedestal is the average response in the absence of deposited energy: its value is very
important because it must be subtracted from
the response of each channel, in order to obtain a response that is directly proportional to the
intensity of the \Laser{} pulses.

Both the pedestal and noise can
be measured in the Pedestals runs and in the Alpha runs. Indeed, in the Alpha runs, for each event,
only one photodiode is hit by an $\alpha$ particle, thus the three other photodiodes contain no
energy; in this case, the pedestal and noise levels can be measured. 
It is also very important for the photodiodes
calibration to make sure that the pedestal and noise are the same in the Pedestals and Alpha runs,
in particular that there is no cross-talk between the channels when some energy is deposited in a
photodiode.

\begin{figure}[t]
  \begin{center}
    \includegraphics[width=0.7\textwidth]{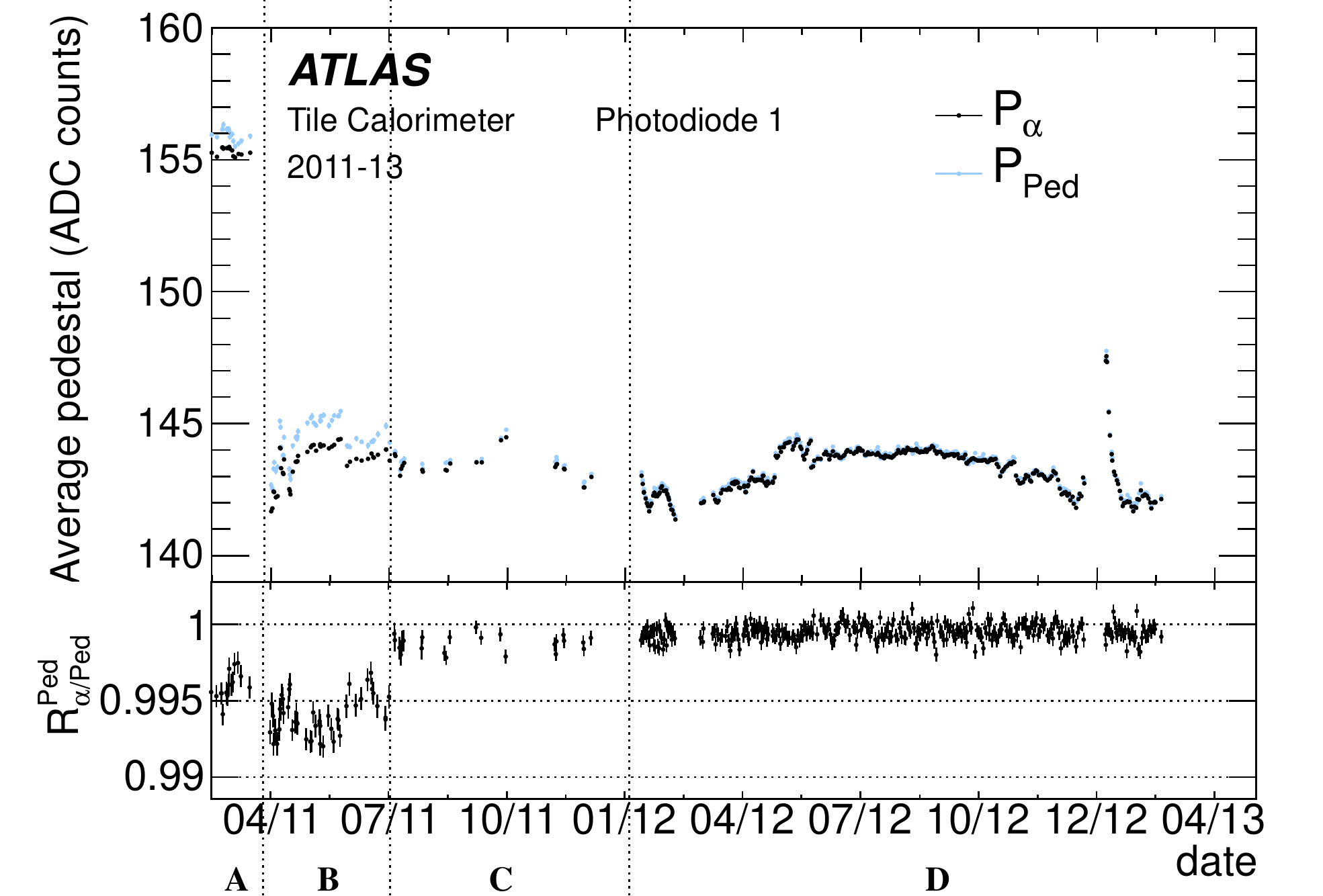}
    \caption{Evolution of the pedestal value measured in the Pedestals runs and in the Alpha 
      runs, and their ratio for the first photodiode from February 2011 to February  2013. 
      The error bars represent the statistical uncertainties on the averages and the 
      ratios. The large variation early 2013 is due to the stabilisation of the readout
      electronics after the end-of-year shutdown, during which all systems were off.}\label{fig:pedestals}
  \end{center}

  \begin{center}
    \includegraphics[width=0.7\textwidth]{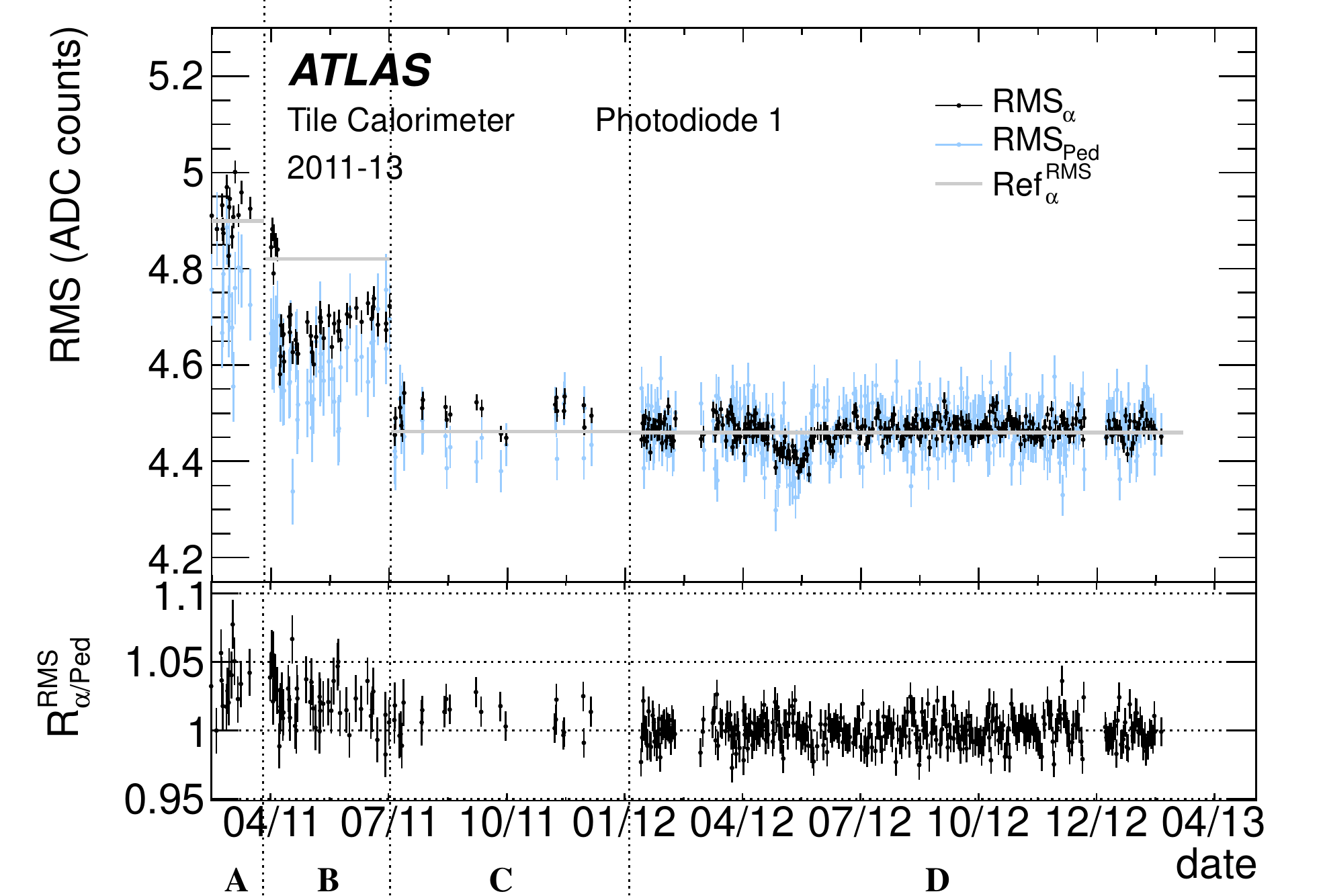}
    \caption{Evolution of the noise value measured in the Pedestals runs and in the Alpha 
      runs, and their ratio for the first photodiode from February 2011 to February  2013. 
      The error bars represent the statistical uncertainties on the averages and the 
      ratios.}\label{fig:noise}
  \end{center}
\end{figure}

The stability of the electronics has been studied during the LHC data taking periods of 2011,
2012 and 2013, from February 2011 to February 2013. 
Figure~\ref{fig:pedestals} shows the evolution of the pedestal measured in the Pedestals runs 
(P$_\mathrm{Ped}$) and in the Alpha runs (P$_\alpha$), together with the ratio 
R$^\mathrm{Ped}_{\alpha/\mathrm{Ped}}=$P$_\alpha$/P$_\mathrm{Ped}$, for photodiode 1 (similar results
are obtained for the three other photodiodes). The vertical dashed lines
separate the two years in four periods, named A to D, that correspond to stable conditions separated
by hardware interventions on the system. In particular, in March 31$^\mathrm{st}$ 2011 (between periods
A and B), a new version of the photodiodes amplifiers was installed, allowing to switch off
photodiodes collecting 
an amount of light sufficient to saturate the electronics and thus generating cross-talk
with the other channels. The second important intervention, in July 6$^\mathrm{th}$ 2011 (between
periods B and C), was a change in the photodiodes amplifiers 
grounding scheme in order to further reduce the cross-talk between
channels. As can be seen in figure~\ref{fig:pedestals}, the pedestals measured in Pedestals runs and
in Alpha runs are similar (the ratio R$^\mathrm{Ped}_{\alpha/\mathrm{Ped}}$ is very close to 1
and is stable) during periods C and D, thus showing that the cross-talk has been reduced to a negligible
value. This means that the pedestal measured in Pedestals runs
can be safely subtracted from the
measured response of the photodiodes in Alpha runs.

Figure~\ref{fig:noise} shows the evolution of the noise measured in the Pedestals runs 
(RMS$_\mathrm{Ped}$)  and in the Alpha runs (RMS$_\alpha$), together with the ratio
R$^\mathrm{RMS}_{\alpha/\mathrm{Ped}}=$RMS$_\alpha$/RMS$_\mathrm{Ped}$. 
The change of grounding scheme in July 2011
significantly improved the electronic noise. Moreover, the noise
is similar in the Pedestals and in the Alpha runs: R$^\mathrm{RMS}_{\alpha/\mathrm{Ped}}\sim1$ with
variations well within $\pm5$~\%.

The linearity of the readout electronics can be measured by injecting a known charge in the photodiodes
amplifiers and varying this charge over the full dynamic range (see figure~\ref{fig:calib_lin}). 
Using dedicated Linearity runs,
the difference between the electronics response and a straight line varies within $\pm1$~\%, with an RMS
of 0.2~\%.

Finally, using standalone \Laser{} runs during which the shutter was closed, it was also possible to
measure the pedestal and noise of the channels connected to the three photodiodes that see the light
after the light splitter, but in a condition similar to a standard \Laser{} run (\Laser{} pump
on and light pulses seen by the photodiode 1). The pedestal variation with respect to standard
Pedestals runs is less than 1.5~\% and the noise differs\footnote{It must be noted that the noise is
not systematically higher during \Laser{} runs, but sometimes higher sometimes lower, as in Alpha runs.} 
by less than 5~\%. This ensures that no significant cross-talk between photodiodes is observed during
\Laser{} runs.

In conclusion, the readout electronics is found to be stable since April 2011. Evolutions of the
pedestal, probably due to environmental effects like temperature and humidity, is not a problem since the
value measured regularly in the Pedestals runs can be used for the response renormalisation in the
Alpha and \Laser{} runs. Moreover the electronic noise is perfectly stable and similar in different 
types of runs.

For the rest of this article, the response of the photodiodes will always be the value after pedestal
subtraction.

\subsection{Calibration of the photodiodes}
In the \Laser{} calibration of the \TileCal{}, the photodiodes are necessary to measure the intensity
of each \Laser{} pulse sent to the calorimeter PMTs, thus requiring that the response of the
photodiodes to a given energy does not vary with time. The monitoring of this potential evolution is
performed by depositing a known amount of energy periodically in the photodiodes, 
i.e. $\alpha$ particles of 5.6~MeV, using the Alpha runs.

\begin{figure}[t]
  \begin{center}
    \includegraphics[width=0.7\textwidth]{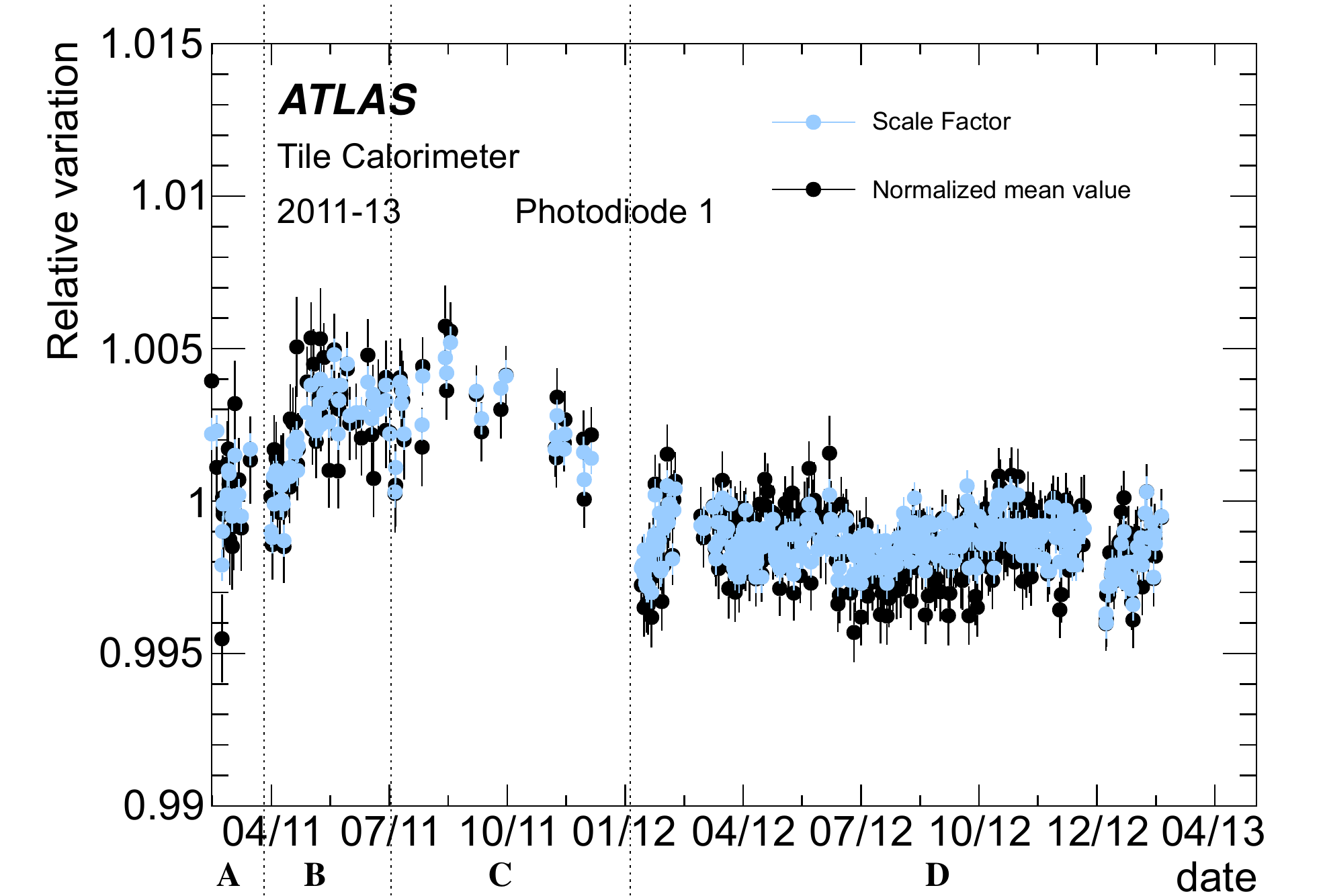}
    \caption{Evolution of the scale factor and the normalised mean value of the $\alpha$ spectra
      from February 2011 to February  2013. 
      The error bars represent the statistical uncertainties.}\label{fig:alphaSF}
  \end{center}
\end{figure}

Two methods have been used to study the evolution of the response of the photodiodes to $\alpha$ 
particles. In both
methods, a high statistics Alpha run is used as a reference and each Alpha run is compared to this
reference. Four reference runs have been defined, each one at the beginning of a period as defined
in the previous section. The reference run for period C contains $3.8\times10^5$ $\alpha$ events and the
reference run for period D $1.1\times10^6$ events.

The first method is using the \emph{normalised mean value} that consists in computing the ratio between
the mean value of the $\alpha$ spectrum in the Alpha run under study with the mean value of the $\alpha$
spectrum in the reference run.

The second method has been developed specifically for this study.
It is called the \emph{scale factor} and it is based on the assumption that the variation in 
the measured $\alpha$ spectra is due to a variation of the gain of the photodiodes and/or their readout 
electronics, thus only implying a rescaling of the spectra and not a distortion. Therefore, in this 
method, the spectrum under study is compared to the reference spectrum. A test spectrum is first built
by rescaling the reference one, i.e. by multiplying the photodiode response for each recorded event by 
a constant number, called the scale factor. Then, this test scale factor is varied until the resulting 
distribution fits as well as possible the $\alpha$ spectrum under study, using the 
Kolmogorov--Smirnov test~\cite{kolmogorov1933,smirnov1948}. The resulting scale factor is 
then a measurement of the gain variation, being equal to one if no rescaling is needed to fit the 
reference distribution to the studied one. 

To compare the two methods and their sensitivity to pedestal and noise variations, studies with
simulated $\alpha$ spectra were performed. These simulated $\alpha$ spectra were generated by
randomly drawing photodiode responses, according to a measured high statistics $\alpha$ distribution.
Then, these simulated $\alpha$ spectra were shifted (to simulate a pedestal variation), rescaled
(to simulate a gain variation) or smeared (to simulate an increase of the noise), before being
used as input to the photodiode calibration procedure.
These studies have shown that
the scale factor method has a smaller uncertainty (0.05~\%) than the normalised mean value (0.1~\%).
Moreover, they have also shown that the bias introduced by a wrong measurement of the pedestal is the
same for the two methods: if the true pedestal differs from the subtracted one by $x$~\%, the bias
is $0.2\times x$~\%.
Finally, as has been shown in the previous
section, the noise in the Alpha runs may be up to 5~\% larger or smaller than in the Pedestals runs.
The simulation showed that an error of 5~\% on the noise would bias the scale factor by a negligible
amount (0.0003~\%) and would have
no measurable effect on the normalised mean value. The noise measured in the reference runs
(Ref$_\alpha^\mathrm{RMS}$) can be seen on figure~\ref{fig:noise} as the horizontal
lines and shows a good agreement with the noise measured in the studied Alpha runs. Therefore, the
scale factor is chosen as the main method to study the evolution of the response of the photodiodes, with the  
normalised mean value as a cross-check.

Figure~\ref{fig:alphaSF} shows the evolution of the normalised mean value and the scale factor 
measured in the Alpha runs. It appears that the period D was more stable than the previous ones.
Concentrating on the scale factor, the largest variation is 0.6~\% in the periods C and D and usually
even less, which is small enough to safely use the response of the photodiodes to \Laser{}
pulses as a reference.

\section{Calibration of the calorimeter}\label{sec:calib}
As has been seen in equation~\ref{eq:E}, the reconstruction of the energy in \TileCal{} depends on
several constants, some of them being updated regularly. The main calibration
of the \TileCal{} energy scale is obtained using the Cesium system~\cite{Aad:2010af}. However,
since a Cesium scan needs a pause in the proton--proton collisions of at least six hours, this
calibration cannot be performed very often. Therefore, regular relative calibrations are accomplished
between two Cesium scans using the \Laser{} system.
The method to compute the \Laser{} constant $\fLas$ introduced in equation~\ref{eq:E} 
is based on the analysis of specific \Laser{} calibration runs, taken about twice a week, 
for which both the \Laser{} system photodiodes and the \TileCal{} PMTs are readout. By definition,
if the response of a channel to a given \Laser{} intensity is stable 
(the gains of the PMT and of the associated readout electronics are stable),
the \Laser{} constant $\fLas$ is 1.

In the next sections, the general method to determine these \Laser{} calibration constants $\fLas$
is first described, the uncertainty on these constants is then estimated,
followed by the description of the actual procedure to produce them.

\subsection{Description of the method}

A \Laser{} calibration consists in two successive runs:
\begin{itemize}
\item a Low Gain run (labelled LG) with $10^4$ pulses using the filter with an
attenuation factor of 3,
\item a High Gain run (labelled HG) with $10^5$ pulses using the filter with an
attenuation factor of 330.
\end{itemize}

For each type of run, $R_{i,p}$ is defined as the normalised response of channel $i$:
\begin{equation}
R_{i,p} = \frac{E^{\rm pmt}_{i,p}}{D1_{p}}
\end{equation}
where, for the pulse $p$, $D1_p$ is the signal measured in the photodiode 1\footnote{The photodiode
1 is the one that measures the light right after the semi-reflecting mirror. The three other 
photodiodes have not been used in this method because several hardware interventions occured during
the two years, affecting the optical fibres connected to these diodes.} in the \Laser{} box and
$E^{\rm pmt}_{i,p}$ is the signal measured by the \TileCal{} photomultiplier connected to the channel $i$.
The analysis is performed with the mean value of this ratio over all pulses for each channel, denoted
$R_{i}=<R_{i,p}>$.

The \Laser{} calibration is a relative calibration with respect to a \Laser{} reference run taken right
after each Cesium scan. The raw relative gain variation of a channel $i$ is defined as follows:
\begin{equation}
\Delta_{i} = \frac{R_{i} - R^{\rm ref}_{i}}{R^{\rm ref}_{i}}
\end{equation}
where $R^{\rm ref}_{i}$ is the normalised response of the \TileCal{} channel $i$ during the reference 
\Laser{} run.

However, due to inhomogeneities in the light mixing of the light splitter 
or radiation damage to the long clear optical 
fibres, the light intensity may vary with time and from fibre to fibre 
(one fibre being linked to half of the PMTs of the same \TileCal{} module).
Therefore, $\Delta_i$ is corrected by the term $\Delta^{\rm fibre}_{f(i)}$ 
($f(i)$ is the number of the fibre coming from the \Laser{} and connected to
the given channel $i$) taking into account the gain variation due to a light 
variation of this specific fibre, as explained in the next paragraph.

\begin{figure}[t]
  \begin{minipage}[t]{0.46\textwidth}
    \includegraphics[width=\columnwidth]{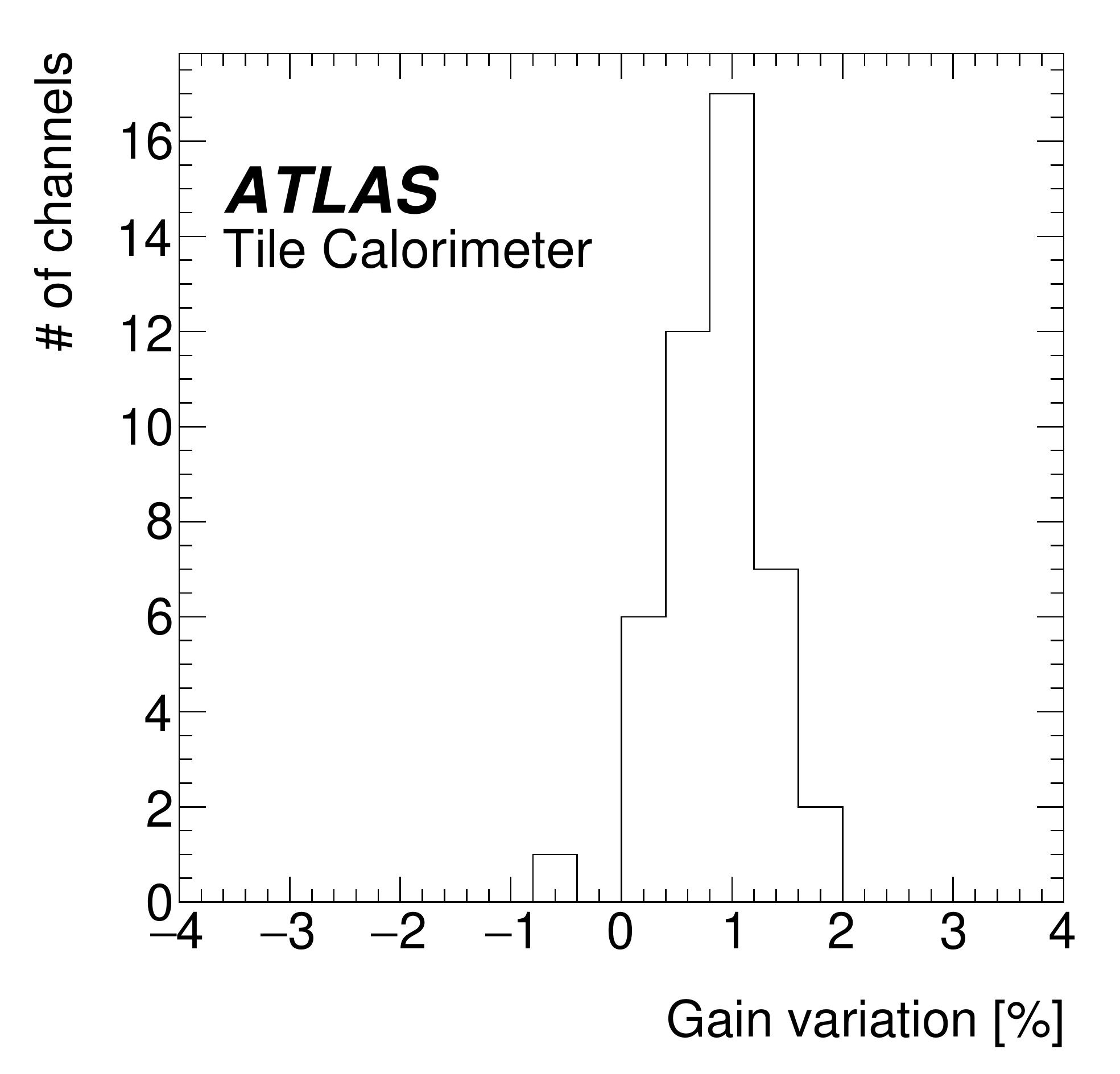}
    \caption{Example of distribution of $\Delta_i$ for all channels connected to one fibre. The 
      average of this distribution is $\Delta^{\rm fibre}_{f(i)}$ (see text).}\label{fig:fibresingle}
  \end{minipage}
  \hspace{0.06\textwidth}
  \begin{minipage}[t]{0.46\textwidth}
    \includegraphics[width=\columnwidth]{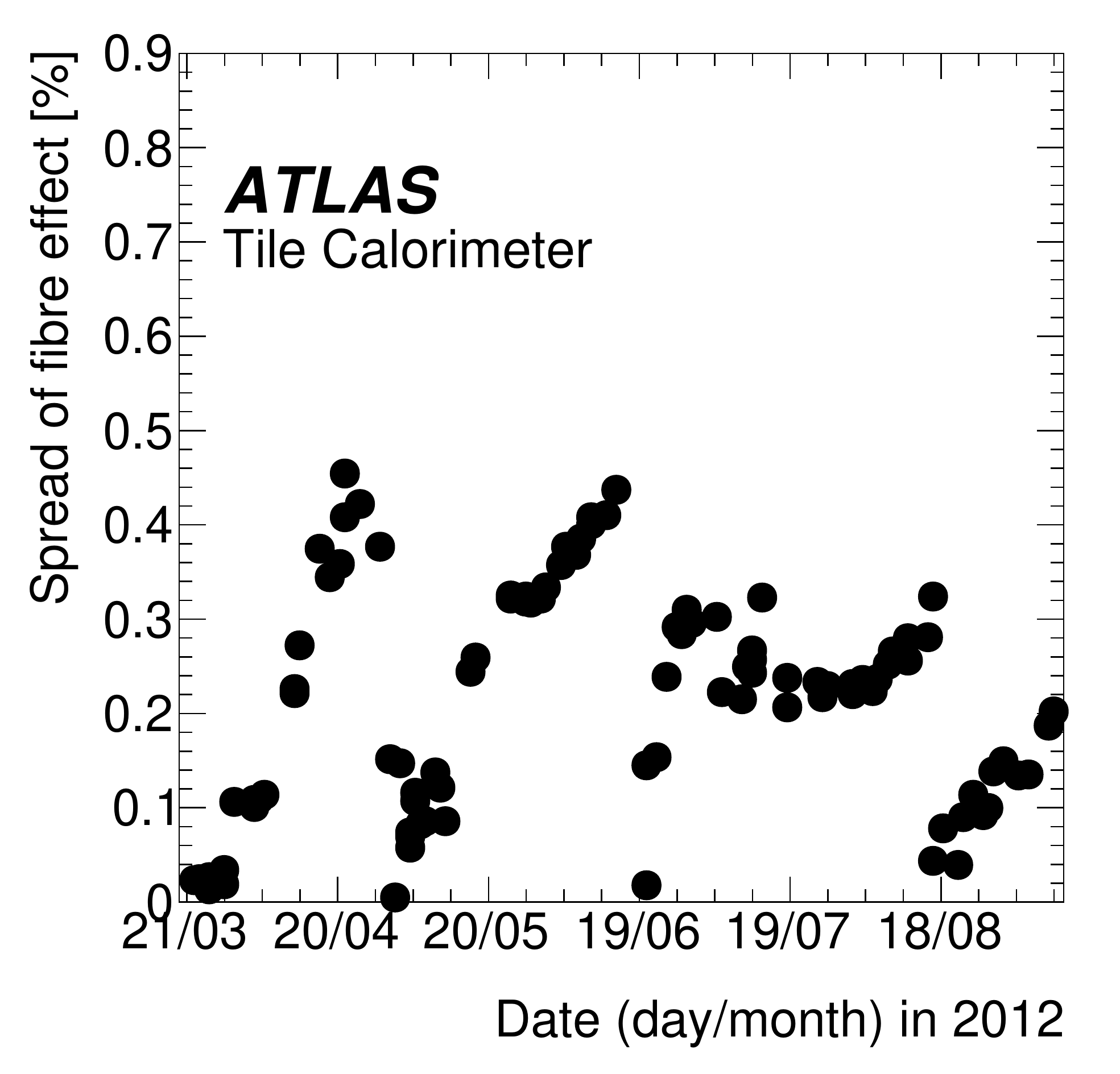}
    \caption{Inhomogeneity due to light distribution instabilities as a function of time.}\label{fig:fibrecorr}
  \end{minipage}
\end{figure}

To compute $\Delta^{\rm fibre}_{f(i)}$, an iterative method is used. First,
the distribution of $\Delta_{i}$ of channels fed by the same fibre is
considered (see figure~\ref{fig:fibresingle}). In this distribution, only PMTs connected
to the D cells for the long barrel modules and the B13, B14, B15, D5 and D6 cells for the
extended barrel modules are used, assuming that these selected
channels are stable between two Cesium scans\footnote{Since the D cells are the most distant cells
from the interaction point, they are less affected by the radiations from the proton--proton collisions, 
unlike A
cells that are more sensitive to the high light dose. The level of radiations has been validated
using the minimum bias current measurements. The evolution of these cells between two Cesium scans
has been checked from the evolution of the Cesium calibration constants and this evolution 
is well below 1~\%.}.
The mean and the RMS of the $\Delta_i$ distribution are then calculated from these
selected channels. To deal with single drifting channels that would bias the measurement of the
shift due to the fibre itself, an
iterative procedure is applied: all the channels that have a gain variation larger
than twice the RMS are excluded from the distribution for the next iteration. The mean gain variation
after five such iterations is the correction $\Delta^{\rm fibre}_{f(i)}$. 



The aim of this method is to correct the drifting channels of \TileCal{}, not the variations of the
fibres or any global variation. Therefore, the 
correction factor, $f^i_{\rm Las}$ for channel $i$, is defined as:
\begin{equation}
f^i_{\rm Las} = \frac{1}{1+\Delta_{i} - \Delta^{\rm fibre}_{f(i)}}
\end{equation}
and is the \Laser{} calibration constant that enters in equation~\ref{eq:E}.

\subsection{Evolution of the fibre correction}


Figure~\ref{fig:fibrecorr} gives a hint of the inhomogeneity of the light distribution during
year 2012, estimated as the RMS of the distribution of all the 
$\Delta^{\rm fibre}_{f(i)}$ with respect to the latest \Laser{} reference run. At the beginning of each 
period, i.e. after each Cesium scan, the spread of the fibre correction factor determination is 0 by definition and it 
increases up to 0.4~\% in a one month period.


\subsection{Uncertainty on the correction factor}
The method assumes that the reference D and B cells are stable.
Systematic effects on the correction factor may arise, in particular from global variations of the
PMT gains of these cells, due to two effects. The first effect, independent of the presence or absence of collisions,
is a constant increase of the gains, but it is negligible over a period of one month (less than 
0.1~$\%$) and has been first observed in 2009 by the Cesium system, prior to collisions.
The second effect, observed during collisions, is a decrease of the gains. 
Its origin is not completely understood but the effect is less than 0.5~$\%$ between two
Cesium scans.


Assuming that the scintillators ageing is negligible between two Cesium scans (typically one month) and taking into account the 
fact that the precision of Cesium calibration constants is at the per-mil-level, the Cesium system is a 
good tool to determine the precision and the systematic uncertainty connected with the \Laser{} 
calibration constants. The two systems are expected to provide compatible measurements. To compare the \Laser{} and the 
Cesium calibration constants, pairs of the closest \Laser{} run/Cesium scan are selected. 


For each pair of \Laser{} run/Cesium scan, the ratio of \Laser{} calibration constants $\fLas$ 
and Cesium calibration constants $\fCs$ is considered. In this comparison, are considered 
only the channels for which the quality of the two calibration constants is good.
The distribution is fitted to a Gaussian function. 
The mean obtained by the fit quantifies the compatibility of the two calibration systems and its difference to one (called $\delta$) is interpreted as the systematic uncertainty on the \Laser{} calibration constants. The $\sigma$ obtained by the fit can be interpreted as the statistical precision of the \Laser{} system assuming that the uncertainty on the Cesium calibration constants is negligible with respect to the \Laser{} one (it is therefore a conservative estimate of the \Laser{} system precision).

\begin{figure}[t]
  \begin{center}
    \subfigure[]{\includegraphics[width=0.49\textwidth]{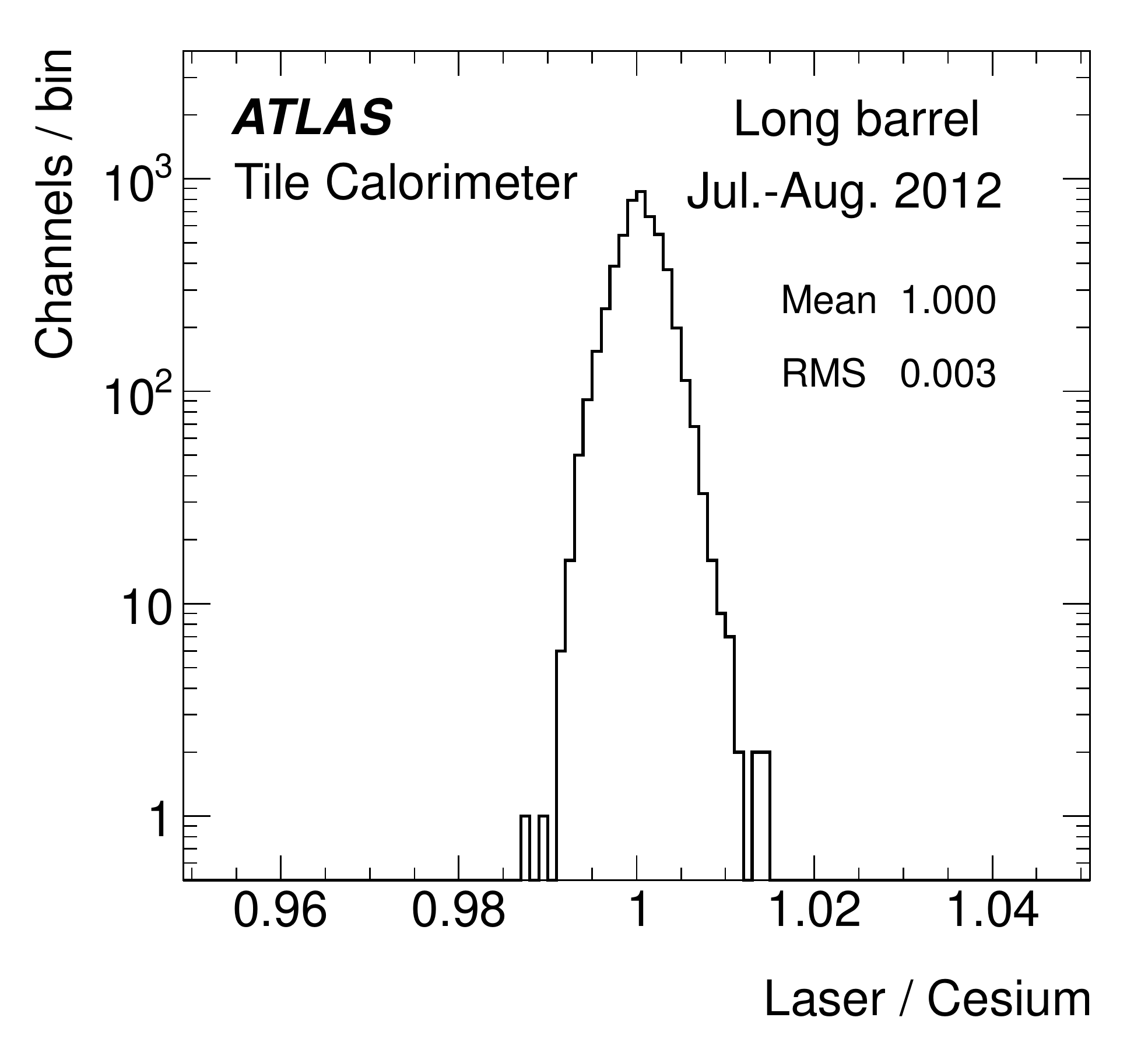}}
    \subfigure[]{\includegraphics[width=0.49\textwidth]{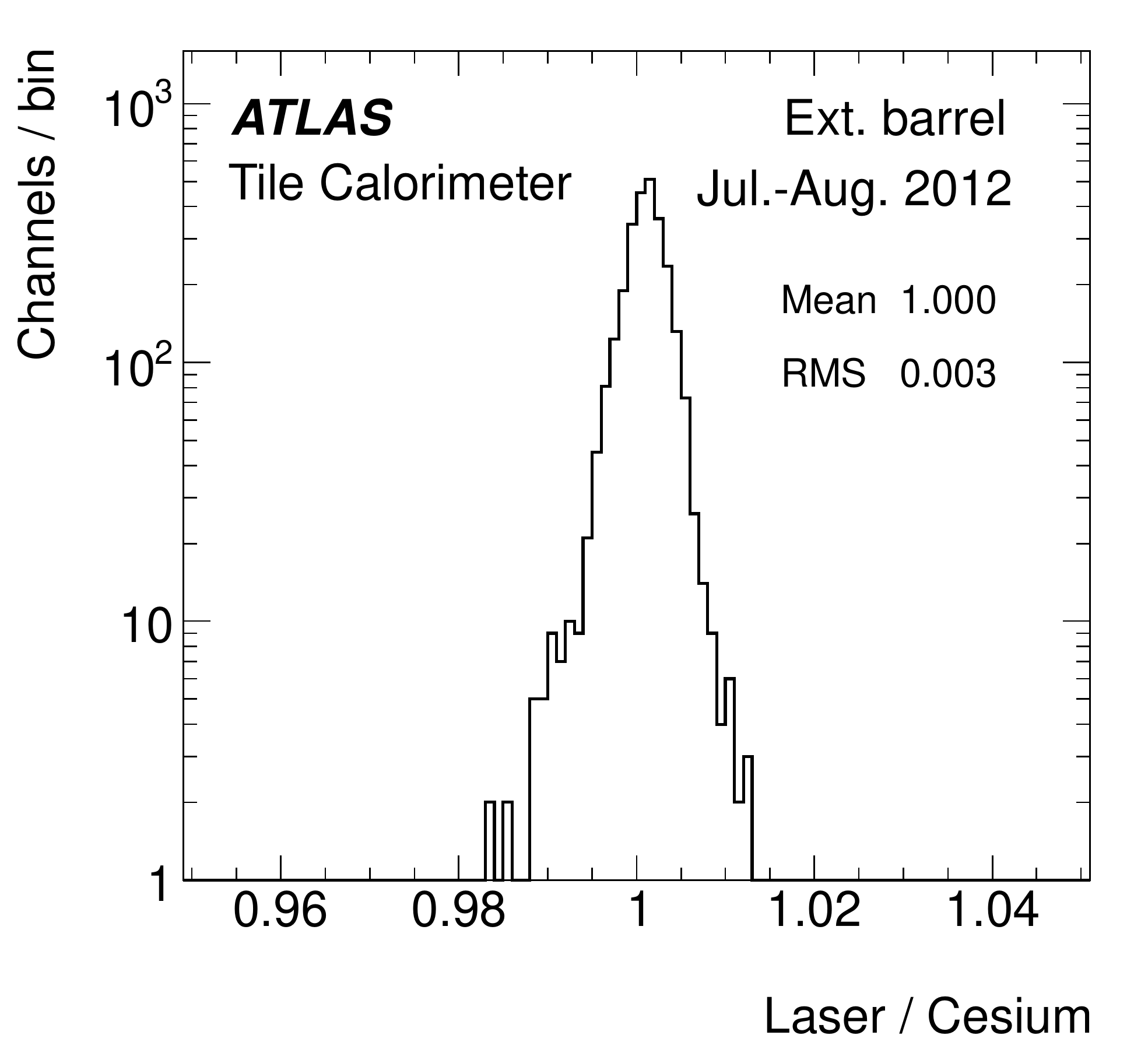}}
    \caption{Example of distribution of the ratio $\fLas$/$\fCs$ during one month for the long barrel
      (a) and the extended barrels (b). The runs were taken in July 9\thi{}
      and August 5\thi{}, 2012.}\label{fig:ratioCsLas}
 \end{center}
\end{figure}

\begin{table}[t]
  \begin{center}
    \caption{Difference between the mean and unity ($\delta$) 
      and spread ($\sigma$) of a Gaussian fit to the distribution of 
      the ratio $\fLas$/$\fCs$ during 2012. The value of $\delta$ is always positive,
      due to the fact that the reference cells, that are assumed to be stable, have
      nonetheless a small decrease of the gain, yielding a systematic effect.}\label{tab:ratioCsLas} 
    \begin{tabular}{c |c c |c c}
      \hline \hline
      & \multicolumn{2}{c}{Long barrel} & \multicolumn{2}{|c}{Extended barrel} \\
      Period & $\delta$ in~$\%$ & $\sigma$ in~$\%$ & $\delta$ in~$\%$ & $\sigma$ in~$\%$ \\ 
      \hline 
      Mar 18-Apr 27 & 0.20 $\pm$ 0.01 & 0.29 & 0.29 $\pm$ 0.01 & 0.29 \\ \hline
      Apr 28-Jun 18 & 0.21 $\pm$ 0.01 & 0.30 & 0.56 $\pm$ 0.01 & 0.36 \\ \hline
      Jun 18-Jul 9 & 0.20 $\pm$ 0.01 & 0.27 & 0.40 $\pm$ 0.01 & 0.27 \\ \hline
      Jul 10-Aug 4 & 0.10 $\pm$ 0.01 & 0.26 & 0.10 $\pm$ 0.01 & 0.24 \\ \hline
      Aug 5-Aug 16 & 0.10 $\pm$ 0.01 & 0.27 & 0.10 $\pm$ 0.01 & 0.25 \\ \hline
      Aug 17-Sep 11 & 0.0 $\pm$ 0.01 & 0.27 & 0.10 $\pm$ 0.01 & 0.25 \\ \hline
      Sep 11-Sep 20 & 0.10 $\pm$ 0.01 & 0.26 & 0.30 $\pm$ 0.01 & 0.22 \\ \hline
      Sep 22-Oct 9 & 0.02 $\pm$ 0.01 & 0.26 & 0.05 $\pm$ 0.01 & 0.24 \\ \hline
      Oct 10-Nov 3 & 0.03 $\pm$ 0.01 & 0.28 & 0.13 $\pm$ 0.01 & 0.29 \\ \hline
      Nov 6-Dec 4 & 0.20 $\pm$ 0.01 & 0.24 & 0.30 $\pm$ 0.01 & 0.25 \\ \hline
      Dec 7-Dec 10 & 0.0 $\pm$ 0.01 & 0.36 & 0.40 $\pm$ 0.01 & 0.46 \\ \hline
      \hline
    \end{tabular}
  \end{center}
\end{table}

Figure~\ref{fig:ratioCsLas} shows a comparison between \Laser{} and Cesium calibration 
constants, the ratio $\fLas$/$\fCs$, for long and extended barrels, for a one month period in
2012.
Fitting the same distributions for 11 periods in 2012 (see results in table~\ref{tab:ratioCsLas}), the systematic uncertainty on $f_{las}$ is evaluated to be at
most 0.21~\% for the long barrel and 0.56~\% for the extended barrels, and the statistical uncertainty of the \Laser{} system varies between 0.22~\% and 0.46~\%. The second period corresponds to a period of high LHC luminosity. As anticipated, this high luminosity has a more significative effect on the PMTs of the extended barrels
than on the ones of the long barrel, as can be seen from table~\ref{tab:ratioCsLas}.
The statistical and systematic uncertainties of the correction factors obtained using the
\Laser{} system are then estimated:

\begin{itemize}
\item statistical uncertainty: 0.3~\%
\item systematic uncertainty for long barrel: about 0.2~\%
\item systematic uncertainty for extended barrels: about 0.5~\%
\end{itemize}

Combining together these results, the overall precision is estimated to be 0.4~\% and 0.6~\% for the long and the extended barrels respectively. 
However, the precision on the Cesium calibration constants (of the order of 0.2~\%) is not completely
negligible with respect to the \Laser{} ones, thus the above values must be considered as upper
limits rather than exact values.

\subsection{Determination of the calibration constants}\label{sect:det_consts}
In equation~\ref{eq:E}, $\fLas$ represents the correction of the PMT gain variation computed with
the \Laser{} system. However, only channels that have undergone a significant deviation are corrected,
i.e. the value of $\fLas$ is set to 1 if it is smaller than a predetermined threshold of 
about three times the overall precision of the \Laser{} system, thus a threshold of 1.5~\% 
for the long barrel and 2~\% for the extended barrels. The low gain (LG) and the high gain (HG) \Laser{} 
runs are used to determine the \Laser{} calibration constants. However, the LG data are more precise than
the HG data because the HG signal amplitude is much smaller than the LG one (a factor 100 between them 
for the same channel). Therefore, the LG \Laser{} calibration constants are used for both gains, while
the HG Laser{} calibration constants are used for cross-checks. A readout electronics issue can be 
revealed by an incompatibility between the LG and the HG \Laser{} calibration constants.

Therefore, the production of the \Laser{} calibration constants $\fLas$ follows this procedure:
\begin{itemize}
\item for each pair of the LG and the HG runs, a \Laser{} calibration constant is computed,
\item a channel is corrected if its LG gain variation is larger than 1.5~\% (2~\%) in the LB (EB), 
\item the compatibility of the \Laser{} calibration constants for the LG and the HG is required
(both constants with same sign and above the thresholds),
otherwise the \Laser{} calibration constants are set to~1 --- this is to ensure the calibration
constant corrects only drifts due to the PMTs and not problems in the readout electronics itself that
are covered by the CIS,
\item a limit is applied to the constants values corresponding to a deviation between
-60~\% and +60~\%. A healthy channel is not supposed to drift up to these values. However, a variation of up to -90~\% is allowed for channels in a module in reduced-HV mode\footnote{During the so-called reduced-HV mode of a module, the High Voltages (HV) regulation loop is turned off and then the HV applied to each PMT is lower, preventing high trip rate. The PMT gain is sensitive to variations of HV, so if the HV of the PMT decreases, the PMT gain will also decrease and will be very low ($\sim$-60~\%) with respect to PMTs supplied with nominal HV.},
\item if a channel is known as bad, its \Laser{} calibration constant is set to 1, whatever is the nature of the problem. Indeed, this flag is very often due to readout electronics, a gain variation or drift faster than \Laser{} run frequency, or corrupted data, and the \Laser{} system is not presumed to correct these cases.
\end{itemize}

Based on these criteria, only 60 channels among $\sim10^4$ could not be corrected by the \Laser{} system during the LHC run 1.

A few additional conditions are needed for three different types of channels: the ones linked to the
E3 and E4 cells, the channels in reduced-HV modules and some channels that have an erratic behaviour,
likely due to readout electronics problems. These latter channels are not calibrated.

The channels linked to E3 and E4 cells are not calibrated by the Cesium
system, while the \Laser{} system monitors all the \TileCal{} channels, including those
connected to these special cells. The fact that there is no Cesium calibration implies that
there are no references for these channels. The chosen solution is to monitor these channels and provide constants with respect to a reference date set to March $18^{\mathrm{th}}$ 2012 which is the date of the first \Laser{} reference run in 2012. Even though the precision of the \Laser{} system
over a period of one year has been estimated to be less than 2~\%, it is sufficient to calibrate 
these highly drifting cells.

Some reduced-HV modules can not be calibrated by the Cesium system. As for the
E3 and E4 cells, these channels are monitored and \Laser{} constants are computed starting from
the \Laser{} reference run taken just before the beginning of the reduced-HV mode period.

Finally, the channels linked to E1, E2 cells\footnote{The E1 and E2 cells are
not calibrated by the \Laser{} because their measured variation, smaller than for the E3 and E4
cells, is of the same order of magnitude as the \Laser{} precision, due to a lower precision
non-standard Cesium calibration (the Cs source cannot pass through these cells).} 
and MBTS (see section~\ref{sect:cells}) are not corrected by the \Laser{} system.



Over the large number of runs recorded and analysed in 2012, only 32 pairs of \Laser{} runs were 
needed to calibrate the data.  Indeed, if several consecutive runs lead to
compatible constants, only the constants from the first pair are implemented for the full period.
The main periods during which a large number of channels had to be calibrated (at most 432 in the
long barrel and 148 in the extended barrels) are at the end of April and the beginning of June 2012.
Indeed, during these periods high luminosity was delivered by LHC, producing high current in the PMTs,
thus modifying their gain. Other channels that needed calibration in 2012 
were the 224 channels reading the E3 and E4 cells as well as the 167 channels of the 
reduced-HV mode modules.

\subsection{Calibration with pulses during physics runs}
Up to now, the \Laser{} calibration constants have been computed using dedicated \Laser{} calibration
runs, recorded on a regular basis between collision runs. But, as explained in 
section~\ref{sect:laser_mode}, the
\Laser{} pulses are also emitted during collision runs, and are used only so far to monitor the calorimeter
timing (see section~\ref{sec:timing}). 
However, with the increase in instantaneous luminosity expected from the
LHC, resulting in possible fast variations of the gains of the \TileCal{} PMTs during collisions, being able to
monitor these gains during the physics runs will become important.
Technically, the main difference with standard \Laser{}
runs is that the number of \Laser{} events is not fixed, since it depends on the length of the 
physics run and therefore on the beam conditions. An additional study was performed and 
showed that, applying the method described in the previous paragraphs, \Laser{}
calibration constants can be derived with a good precision using these runs: 
the statistical precision with 5000 pulses, corresponding to a rather short physics run (85 minutes), is similar to a standard calibration run (0.3~\%).
Therefore, the physics runs will probably be used in the future to derive \Laser{} calibration 
constants.


\section{Timing monitoring and correction}\label{sec:timing}
The \TileCal{} does not only provide a measurement of the energy that is deposited in the calorimeter,
but also the time when this energy was deposited. This information is in particular exploited in the removal of signals that do not originate from proton--proton collisions 
as well as for time-of-flight measurements of hypothetical heavy slow
particles that would enter in the calorimeter. Therefore, the time synchronisation of all calorimeter 
channels represents one of the important issues in the calibration chain.

\subsection{Time reconstruction}
As described in section~\ref{sect:readout}, 
the signal coming from the PMTs is digitised every 25~ns and seven samples
are readout after a positive level-1 trigger decision. 
The optimal filtering (OF) method that is applied on these samples is able to determine the
total amplitude of the signal but also the signal phase, 
defined as the phase of the analog signal peak with respect to the TTC clock signal
corresponding to the fourth sample. The precision on this measurement is better than 
1~ns~\cite{Adragna2009362,Aad:2010af}.
The optimal filtering method is applied online to all channels and later re-computed offline in 
channels that have sufficiently large signals.

However, the parameters of the OF
method depend on this signal phase, thus a bad setting of the timing of a channel will produce
a wrong measurement of the signal amplitude. Studies have shown that a difference of 25~ns between
the actual and supposed phases degrades the reconstructed energy by 35~\%. For
timing differences of a few nano-seconds, the effect on the measured energy is small but the 
corresponding cells could be excluded from the reconstruction, being falsely tagged as not coming
from proton--proton collisions.

The PMT signal digitisation is performed by an electronic board called the digitiser that is able
to process the signal from six PMTs. Only the timing of each digitiser can be adjusted independently of each other. Inside the digitiser the six channels have the same relative timing.

The meaurement of the pulse timing is made in two steps. First, the phases of the digitisers
are set so that the physics signals from particles
originated in the interaction point and travelling with the speed of
light peak close to the fourth sample. Second, the residual offsets
t$_{\mathrm{phase}}$ are measured, using dedicated runs, as the average of the
reconstructed OF time t$_{\mathrm{OF}}$.
The channel time is then defined as
t$_{\mathrm{channel}}^{\mathrm{phys}} =$t$_{\mathrm{OF}} -$t$_{\mathrm{phase}}$. This value should of course be
close to zero for standard particles coming from the interaction point at nearly the speed of light.

\subsection{Timing jumps}
\begin{figure}[t]
  \begin{minipage}[t]{0.46\textwidth}
    \includegraphics[width=\textwidth]{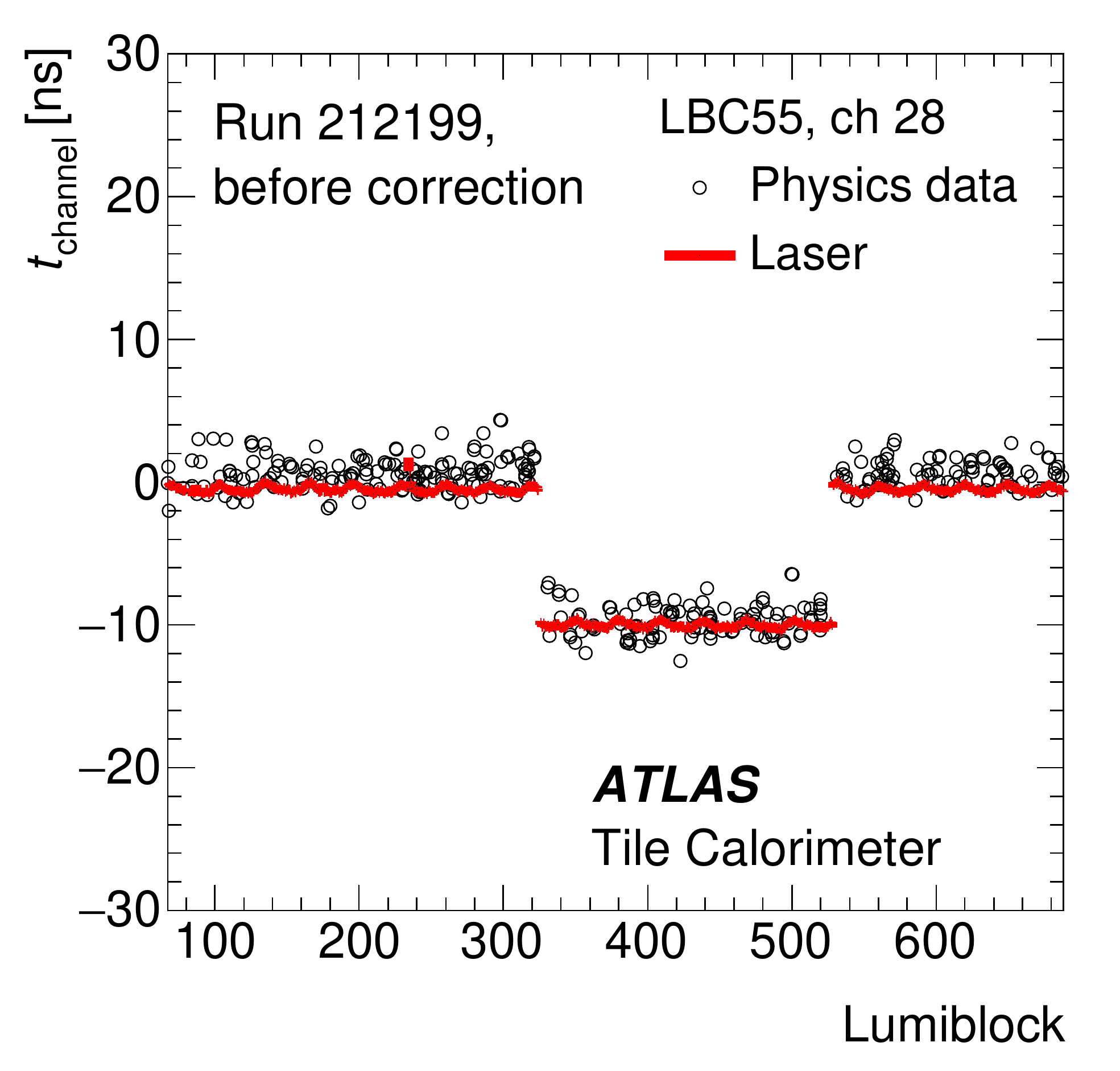}
    \caption{Example of measured channel time as a function of time during a single \Atlas{} run,
    comparing timing from physics events and from \Laser{} calibration events, exhibiting a timing
    jump.}\label{fig:timingjumpsa}
  \end{minipage}
  \hspace{0.06\textwidth}
  \begin{minipage}[t]{0.46\textwidth}
    \includegraphics[width=\textwidth]{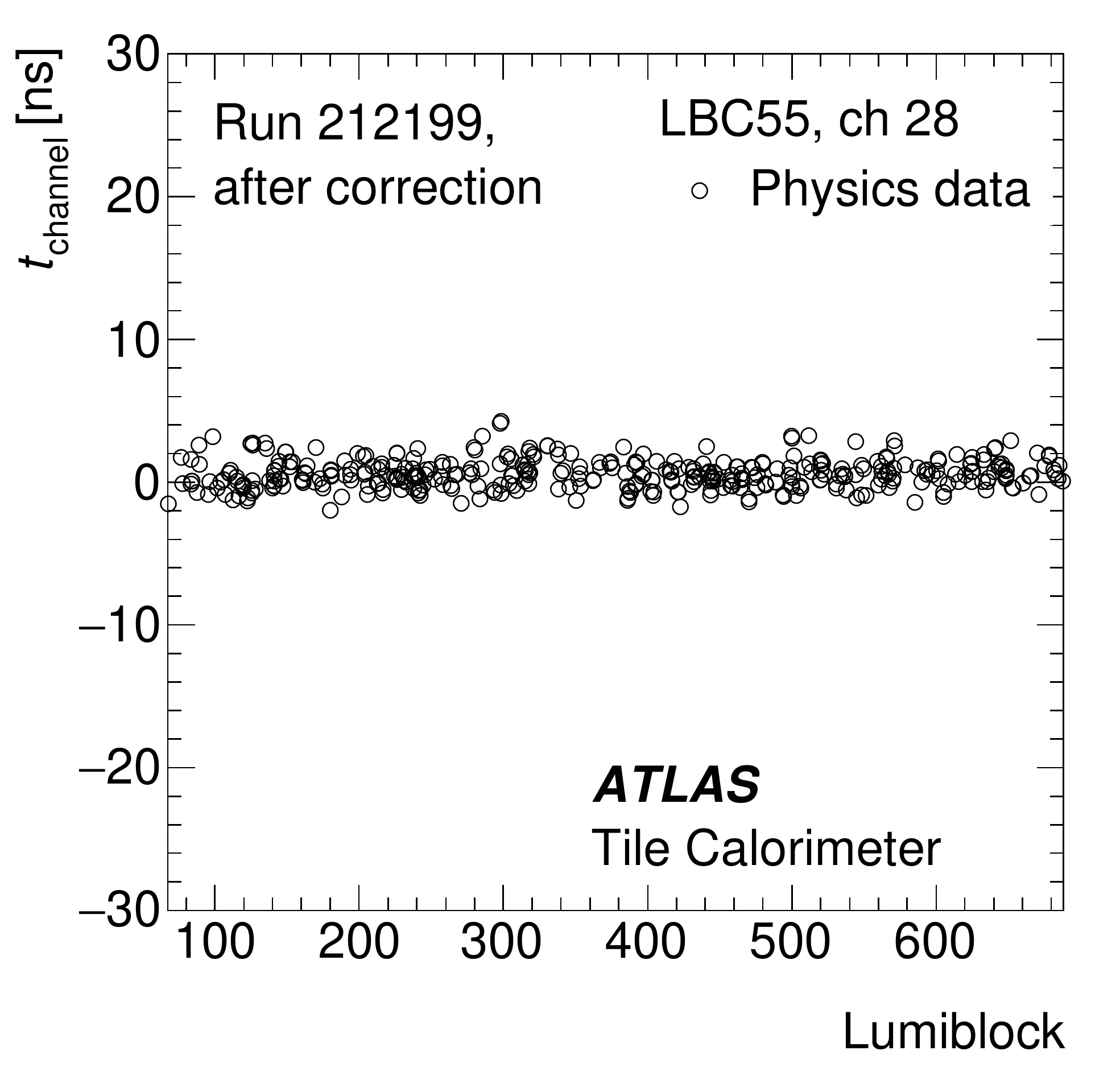}
    \caption{Example of reconstructed time from physics events after correcting for the
    timing jump.}\label{fig:timingjumpsb}
  \end{minipage}
\end{figure}

During data taking, it was discovered that digitisers can suddenly change their timing settings due
to a mis-configuration, either at the beginning of a run or after an automatic power cycling of the 
low voltage power supplies 
(LVPS) feeding the \TileCal{} front-end electronics. This mis-configuration of the time settings
is later referred to as timing jump.
An example of such
a behaviour can be seen in figure~\ref{fig:timingjumpsa} where the reconstructed time for a
channel is represented as a function of the luminosity block, 
that is the elementary unit of time in a run
(one or two minutes depending on data taking periods)
for which parameters like calibration constants can be modified if necessary. 
Of course, this kind of feature has
a negative impact on the calorimeter performance since the reconstructed energy and time are
mis-measured during the period affected by the timing jump.


As stated previously, the \Laser{} system emits light pulses also during physics runs.
Since all calorimeter channels are
exposed to \Laser{} light at the same time, the statistics is sufficient
to determine the reconstructed time to a precision of about
1~ns. Because the \Laser{} is not synchronised with the 
LHC clock, the reconstructed time in each channel must be corrected for the
\Laser{} phase using the \Laser{} TDC information t$_{\mathrm{TDC}}$. 
In addition, the time of arrival of the
\Laser{} pulses to the \TileCal{} PMTs is different from the signal time for particles from 
collisions, due to the path followed by the \Laser{} light. 
Therefore, in a channel perfectly timed-in for physics, the
\Laser{} pulse arrives at a different time\footnote{The
reference \Laser{} time t$^{\mathrm{laser}}_{\mathrm{ref}}$ represents the mean value of
t$_{\mathrm{OF}}$ corrected for the \Laser{} TDC on the per-event basis.} t$^{\mathrm{laser}}_{\mathrm{ref}}$.
The channel time for \Laser{} events is then defined as 
t$_{\mathrm{channel}}^{\mathrm{laser}} =$t$_{\mathrm{OF}} -$t$_{\mathrm{TDC}} -$t$^{\mathrm{laser}}_{\mathrm{ref}}$.
If the time setting of the digitiser is correct, t$_{\mathrm{channel}}^{\mathrm{laser}}$ should be
close to zero.

The \Laser{} events are recorded in parallel to the physics data taking
and this stream gets reconstructed immediately once the run
finishes. The reconstructed \Laser{} times t$_{\mathrm{channel}}^{\mathrm{laser}}$ are
histogrammed for each channel as a function of the luminosity
block. 
Since this time is supposed to
be close to zero, the monitoring program searches for differences from
this baseline. Identified cases --- potential timing jumps --- are
automatically reported to the data quality team for manual
inspection. If the timing jump is confirmed, the t$_{\mathrm{phase}}$ value of the
corresponding channel is modified for the affected period, thus allowing to
recover a correct reconstructed time, as demonstrated on figure~\ref{fig:timingjumpsb}.
These \Laser{} results are available still during the calibration loop
and thus allow for time constants correction before the full data
processing starts.

Almost all timing jumps detected with \Laser{} events can be corrected
during the subsequent data processing. 
However, few digitisers appeared to be very unstable, exhibiting very large number of
timing jumps. Such channels are flagged as \emph{bad timing}.
This flag prevents the time
from such channels to be used in the further physics object reconstruction.

\subsection{Impact of the timing jumps on the calorimeter time performance}
The overall impact of the timing jump corrections on the reconstructed
time was studied with jet collision data using a sub-set of the 2012 data
representing $1.3~\mathrm{fb}^{-1}$ of collected physics data. To
reduce the impact of the time
dependence on the reconstructed energy, the channel energy was
required $E_{\mathrm{channel}} > 4$~GeV, but still read out with
high gain.\footnote{This is important since the high gain response is
  delayed with respect to that of low gain by about 2~ns. The
  high gain requirement imposes an effective upper limit of about
  $E_{\mathrm{channel}} < 12$~GeV.}

\begin{figure}[t]
  \begin{center}
    \includegraphics[width=0.6\textwidth]{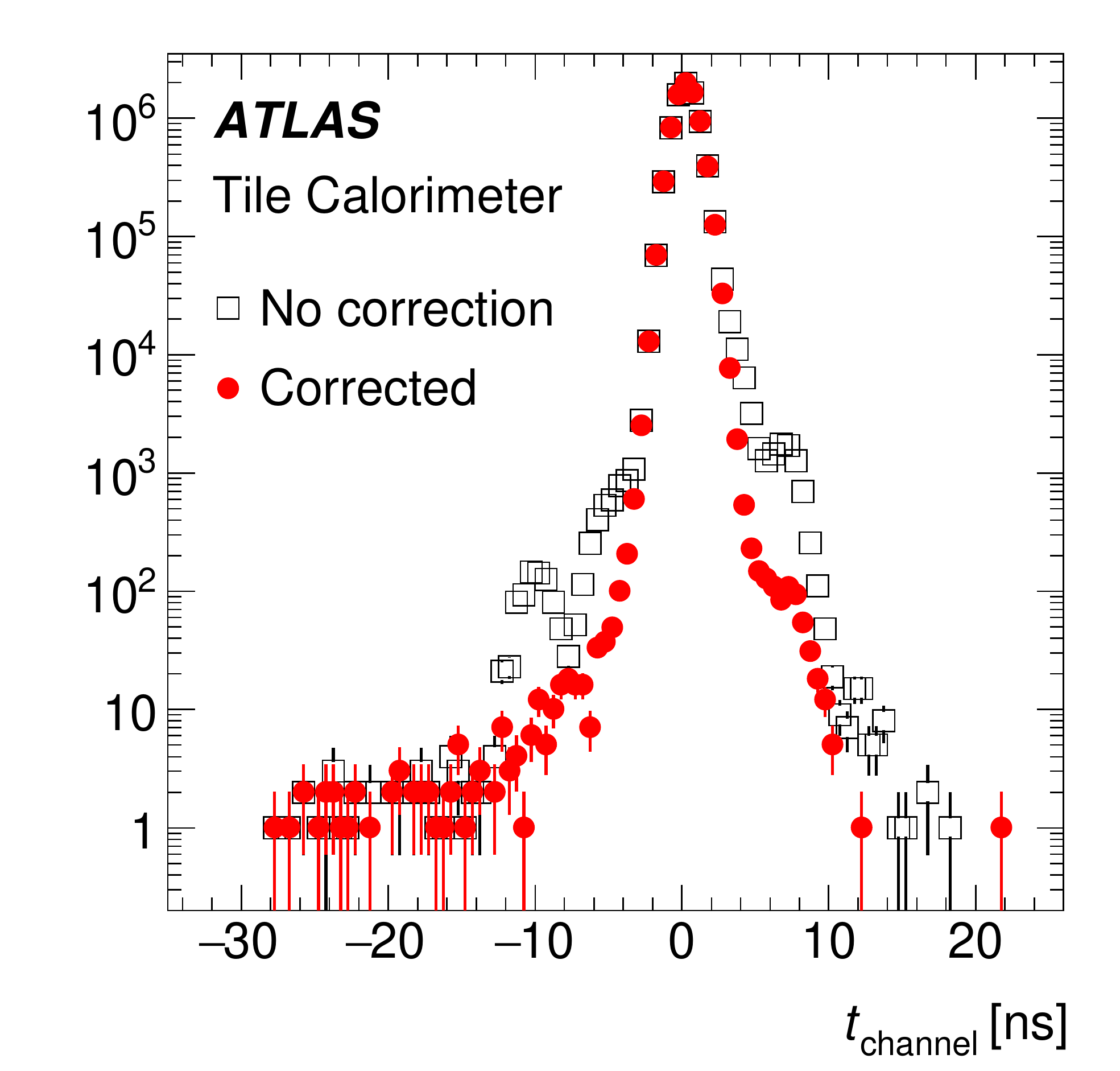}
    \caption{The impact of the timing jump corrections on the
    reconstructed channel time in jet collision data. Shown are all \TileCal{}
    high gain channels with $E_{\mathrm{channel}} > 4$~GeV, being a
    part of reconstructed jets. Data correspond to $1.3~\mathrm{fb}^{-1}$ of collected 
    physics data in 2012.}\label{fig:tchannelG}
 \end{center}
\end{figure}

The results are displayed in figure~\ref{fig:tchannelG}, where
the reconstructed time is shown for all calorimeter channels with and
without the timing jump corrections. While the Gaussian core,
corresponding to channels (and events) not affected by timing jumps, remains
basically unchanged, the timing jump corrections significantly reduce
the number of events in the tails. The overall RMS is improved by 9\,\% (from
0.90~ns to 0.82~ns after the corrections are applied).

\section{Pathological channels monitoring}\label{sec:monitor}

\begin{figure}[t]
  \begin{center}
    \subfigure[\label{fig:laser_HV:OK}]{\includegraphics[width=0.49\textwidth]{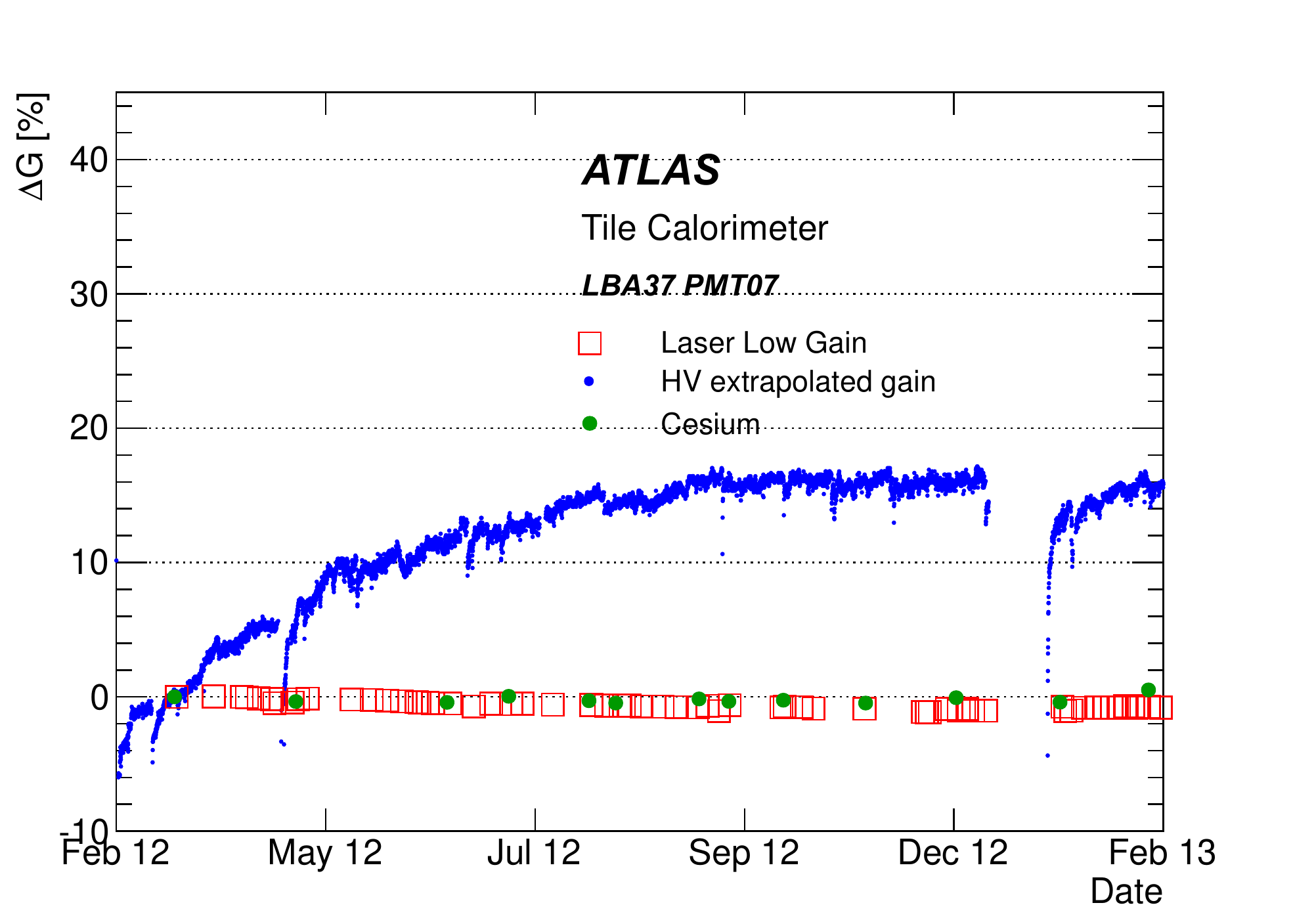}}
    \subfigure[\label{fig:laser_HV:bad}]{\includegraphics[width=0.49\textwidth]{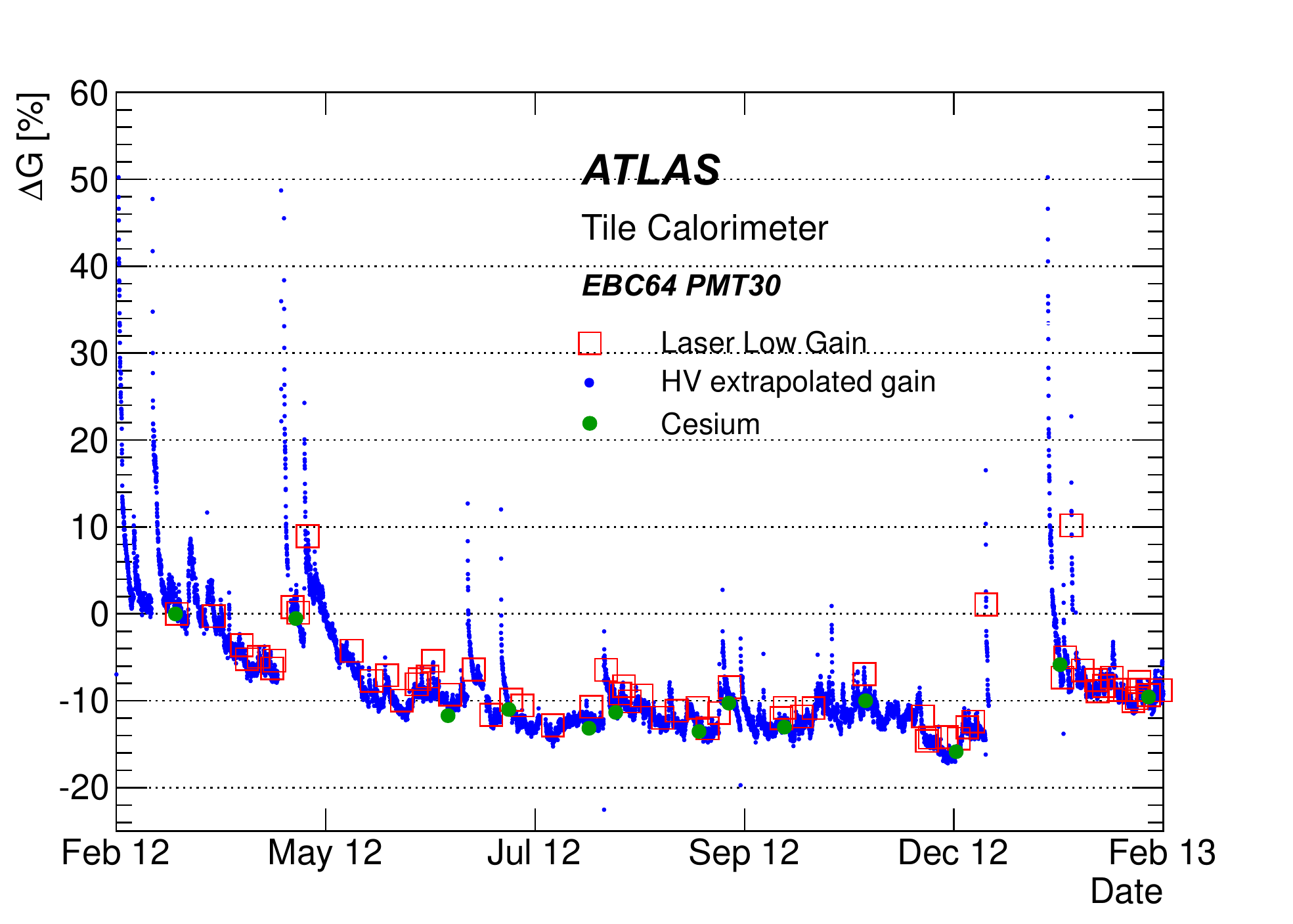}}
    \caption{Example of pathological channels monitoring in 2012. These plots present 
      the comparison between the gain variation 
      expected from the high voltage monitoring (blue dots) and the one measured 
      by the \Laser{} (red squares) and the Cesium (green dots) systems. 
      The vertical structures are due to ON/OFF switchings and are expected.
      However, between two switchings and for normal channels, the gain should be constant.}\label{fig:laser_HV}
 \end{center}
\end{figure}

For a few channels, the
monitoring shows that the PMT input high voltage is unstable. This instability may be due to a true
unstable input voltage or a failure of the monitoring system. In order to disentangle the two
possibilities, the \Laser{} system can be used to measure the real gain variation (as described in the previous
sections).
Figure~\ref{fig:laser_HV:OK} presents an example of such pathological channel behaviour. In this case, the
\Laser{} system does not observe any variation of the PMT gain, meaning that the actual input high voltage
must be stable and the drift observed in the monitoring is not real.
In the case of figure~\ref{fig:laser_HV:bad}, the \Laser{} system confirms
the instability of the high voltage distribution system, the very large gain variations measured by the \Laser{}
system being in agreement with the variations computed from the high voltage monitoring.

\section{Conclusions}\label{sec:concl}
The \Laser{} system is one of the three calibration systems of the \Atlas{} Tile Calorimeter. It is a key
component for monitoring and calibrating the 9852 \TileCal{} photomultipliers and their associated 
readout electronics. The \Laser{} calibration constant $\fLas$ applied on each channel has been regularly
measured, with a statistical uncertainty at the level of 0.3~\%, and is one ingredient in the \TileCal{} 
cell energy reconstruction. Comparing these correction factors with the ones derived with the Cesium 
calibration system, a good compatibility between them was found and the systematic uncertainty
on the \Laser{} calibration constants is estimated to be 0.2~\% for the long barrel modules and 0.5~\% for the
extended barrel modules. The \Laser{} system has also been used to monitor the timing of the
\TileCal{} front-end electronics during physics runs, allowing to correct for timing jumps provoked by
mis-configuration resulting from power supplies trips.
It has also been used to monitor the behaviour of pathological calorimeter channels.

In order to improve the stability and the reliability of the system for the LHC run 2, a major 
upgrade was performed in October 2014, including a completely new electronics, ten photodiodes
to measure the light at various stages in the \Laser{} box, and a new light splitter.

\bibliographystyle{JHEP} 
\bibliography{Laser}

\newpage
{\bf \Atlas{} Tile Calorimeter system}
\\

\noindent
J.~Abdallah$^{20}$,
C.~Alexa$^{29}$,
Y.~Amaral~Coutinho$^{37}$,
S.P.~Amor~Dos~Santos$^{8,26}$,
K.J.~Anderson$^{14}$,
G.~Arabidze$^{5}$,
J.P.~Araque$^{26}$,
A.~Artamonov$^{21}$,
L.~Asquith$^{18}$,
R.~Astalos$^{17}$
% ...
\\

\noindent$^{\dagger}$ deceased\\
% ...
$^{5}$ Department of Physics and Astronomy, Michigan State University, East Lansing MI, United States of America\\
% ...
$^{8}$ Department of Physics, University of Coimbra, Coimbra, Portugal\\
% ...
$^{14}$ Enrico Fermi Institute, University of Chicago, Chicago IL, United States of America\\
% ...
$^{17}$ Faculty of Mathematics, Physics \& Informatics, Comenius University, Bratislava, Slovak Republic\\
$^{18}$ High Energy Physics Division, Argonne National Laboratory, Argonne IL, United States of America\\
% ...
$^{20}$ Institut de F{\'\i}sica d'Altes Energies (IFAE), The Barcelona Institute of Science and Technology, Barcelona, Spain\\
$^{21}$ Institute for Theoretical and Experimental Physics (ITEP), Moscow, Russia\\
% ...
$^{26}$ Laborat\'orio de Instrumenta\c{c}\~ao e F\'\i sica Experimental de Part\'\i culas - LIP, Lisboa, Portugal\\
% ...
$^{29}$ National Institute of Physics and Nuclear Engineering, Bucharest, Romania\\
% ...
$^{37}$ Universidade Federal do Rio De Janeiro COPPE/EE/IF, Rio de Janeiro, Brazil\\
% ...

\end{document}